\RequirePackage{ifpdf}
\documentclass[letterpaper]{article} 
\usepackage[usenames,dvipsnames]{pstricks}
\usepackage{./auxi/jheppub}
\usepackage[english]{babel}
\usepackage[babel]{csquotes}
\usepackage{amsmath}
\usepackage{amsfonts}
\usepackage{amssymb}
\usepackage{mathtools,colonequals}
\usepackage{mathrsfs}
\usepackage{bm}
\usepackage{bbold}
\usepackage{braket}
\usepackage{graphicx}
\usepackage{array}
\usepackage{color}
\usepackage{subcaption}

\DeclarePairedDelimiter\floor{\lfloor}{\rfloor}

\newcommand{\al}{\alpha}

\newcommand{\ga}{\gamma}
\newcommand{\de}{\delta}
\newcommand{\D}{\Delta}

\renewcommand{\th}{\theta}

\newcommand{\La}{\Lambda}
\newcommand{\Ga}{\Gamma}
\newcommand{\m}{\mu}
\newcommand{\n}{\nu}

\newcommand{\si}{\sigma}

\newcommand{\p}{\phi}

\newcommand{\pa}{\partial}
\DeclarePairedDelimiter{\abs}{\lvert}{\rvert}

\newcommand{\beq}{\begin{equation}}
\newcommand{\eeq}{\end{equation}}

\newcommand{\wh}[1]{\widehat{#1}}

\newcommand{\fy}{\widehat{f}}

\newcommand{\bz}{\bar{z}}
\newcommand{\ttr}{\widehat{\tau}}

\def\NN{\mathcal{N}}
\def\BB{\mathcal{B}}
\def\CC{\mathcal{C}}

\newcommand{\Disc}{\textup{Disc}}

\newcommand{\ii}{\mathrm{i}}
\newcommand{\ie}{\emph{i.e.}}

\def\OO{\mathcal{O}}
\def\DD{\mathcal{D}}
\def\CC{\mathcal{C}}
\def\Casdef{\CC_{\mathrm{def}}}
\def\SLDr{\DD_{\parallel}}
\def\Dhat{{\wh{\D}}}
\def\xper{x^{\perp}}
\def\xpar{x^{\parallel}}

\def\Cm{{\mathcal{C}}}

\def\Im{{\mathcal{I}}}

\def\Hm{{\mathcal{H}}}

\def\D{\Delta}

\def\eg{{\emph{e.g.}}}

\newcommand{\rhob}{\bar{\rho}}

\author[a]{Madalena Lemos,}
\author[a]{Pedro Liendo,}
\author[b]{Marco Meineri,}
\author[c,d]{Sourav Sarkar}

\affiliation[a]{DESY Hamburg, Theory Group, Notkestra{\ss}e 85, D-22607 Hamburg, Germany}
\affiliation[b]{Institute of Physics, \'{E}cole Polytechnique F\'{e}d\'{e}rale de Lausanne (EPFL),
Rte de la Sorge, BSP 728, CH-1015 Lausanne, Switzerland}
\affiliation[c]{Institut f{\"u}r Physik und Institut f{\"u}r Mathematik, Humboldt-Universit{\"a}t zu Berlin, IRIS Adlershof, Zum Gro{\ss}en Windkanal 6, 12489 Berlin, Germany}
\affiliation[d]{Max-Planck-Institut f{\"u}r Gravitationsphysik, Albert-Einstein-Institut, Am M{\"u}hlenberg 1, 14476 Potsdam, Germany}

\emailAdd{madalena.lemos@desy.de}
\emailAdd{pedro.liendo@desy.de}
\emailAdd{marco.meineri@epfl.ch}
\emailAdd{sarkar@physik.hu-berlin.de}

\preprint{\parbox{3cm}{DESY 17-239\\ HU-EP-17/31}}

\bigskip
\abstract{
We study the spectrum of local operators living on a defect in a generic conformal field theory, and their coupling to the local bulk operators. We establish the existence of universal accumulation points in the spectrum at large $s$, $s$ being the charge of the operators under rotations in the space transverse to the defect. Our tools include a formula that inverts the bulk to defect OPE, analogous to the Caron-Huot formula for the four-point function \cite{Caron-Huot:2017vep}. Analyticity of the formula in $s$ implies that the scaling dimensions of the defect operators are aligned in Regge trajectories $\Dhat(s)$. These results require the correlator of two local operators and the defect to be bounded in a certain region, a condition that we do not prove in general. We check our conclusions against examples in perturbation theory and holography, and we make specific predictions concerning the spectrum of defect operators on Wilson lines. We also give an interpretation of the large $s$ spectrum in the spirit of the work of Alday and Maldacena \cite{Alday:2007mf}.}

\title{Universality at large transverse spin in defect CFT}


\begin{document}

\maketitle

\section{Introduction and summary}
\label{sec:intro}

The space of conformal field theories (CFTs) is large and varied. Despite this, remarkable features are common to all members of this special family of quantum field theories. The most obvious ones are simple consequences of symmetry, unitarity and locality, which respectively impose that operators form representations of the conformal group, obey unitarity bounds, and that a stress tensor exists. In recent years, it has become clear that universality also emerges at a more detailed level. It was first discovered that the spectrum and OPE coefficients of operators with asymptotically large spin ($\ell$) is tied to the low-lying (low-twist $\tau=\Delta-\ell$) CFT data \cite{Alday:2007mf,Komargodski:2012ek, Fitzpatrick:2012yx}. Recently \cite{Caron-Huot:2017vep}, it was shown that the relation is analytic rather than asymptotic, and that operators in every CFT are organized in Regge trajectories. These facts are best understood as consequences of crossing symmetry, and as such are still rooted in the presence of conformal invariance and unitarity constraints, albeit in a far less obvious fashion. For instance, operators of spin as low as two lie in Regge trajectories, and this is a consequence of boundedness of the four-point function in the so called Regge limit, which in turn follows from reality of the OPE coefficients, \ie\, from unitarity. The contribution of individual Regge trajectories to the crossing equation is enhanced and isolated when pairs of operators become light-like separated. Hence, the lightcone bootstrap is the main source of analytic information on the \emph{double-twist operators} that populate the trajectories.
This line of development nicely complements the study of crossing in Euclidean configuration, which is most sensitive to low-dimensional data, rather than to operators with low twist.
Interestingly, the numerical results obtained by studying the crossing equation around a Euclidean point \cite{arXiv:0807.0004}  (see \cite{Rychkov:2016iqz,Simmons-Duffin:2016gjk} for pedagogical reviews and further references) are also sensitive to the Lorentzian physics of the lightcone limit \cite{Simmons-Duffin:2016wlq}.\footnote{In \cite{Mazac:2016qev} a new analytic approach has been put forward, which provides the exact spectrum exchanged in a four-point function which is extremal in the standard bootstrap sense \cite{ElShowk:2012hu}.}
The combination of the two approaches yields unprecedented amount of information on a non-supersymmetric, higher dimensional CFT such as the critical 3d Ising model \cite{Simmons-Duffin:2016wlq}. 

The spectrum of local operators does not exhaust the set of observables in a CFT. Extended probes arise naturally both from an experimental and from a theoretical point of view. Boundaries and interfaces, surface operators, Wilson and 't Hooft lines are instances of what we refer to collectively as defects. If the dynamics on a defect preserves the spacetime symmetries that do not deform it, we call it a \emph{conformal defect}. Conformal defects support a spectrum of local excitations which are organized according to representations of the preserved symmetries. Locality implies that the spectrum can be studied in the highly symmetric case of a flat or spherical defect, in which the algebra of preserved spacetime transformations is promoted to a full symmetry group.
For reasons to become clear shortly, we are interested in a space-like $p$-dimensional defect in $d$ dimensions, so that the group is $SO(p+1,1)\times SO(d-p-1,1)$. We shall also often denote the codimension as $q$, that is, $p+q=d$. The defect CFT data is constrained by crossing symmetry of the four-point function of defect primaries, and is tied to the bulk via crossing symmetry of correlators involving at least two bulk primaries  \cite{McAvity:1995zd,Billo:2016cpy}. There is a growing effort in refining our understanding of the constraints \cite{Gadde:2016fbj,Hogervorst:2017kbj,Rastelli:2017ecj,Fukuda:2017cup,Lauria:2017}, extracting numerical and analytic information from them \cite{Liendo:2012hy,Gaiotto:2013nva,Gliozzi:2015qsa,Liendo:2016ymz,Gliozzi:2016cmg}, and performing direct computations of correlators in specific models, see \eg, \cite{Billo:2013jda,Cosme:2015cxa,Bianchi:2016xvf,Chiodaroli:2016jod,deLeeuw:2017dkd,Giombi:2017cqn,Soderberg:2017oaa,Melby-Thompson:2017aip,deLeeuw:2017cop,Kim:2017sju}.

In this paper, we address the question whether the defect spectra exhibit universal features akin those briefly discussed above. Of course, crossing symmetry of the defect four-point function implies the existence of double twist defect operators organized in Regge trajectories. Furthermore, the theory on the defect comes naturally endowed with a global symmetry, the $SO(q-1,1)$ group of boosts and rotations around a defect of codimension $q$. We call \emph{transverse spin}, and we denote as $s$, the associated charge. This global symmetry is the main ingredient of our analysis. In what follows, we show that the spectrum of any defect CFT includes universal accumulation points at large $s$. Given any scalar bulk primary operator of dimension $\Delta_\phi$, the defect spectrum contains primaries of dimension 
\beq
\wh{\D}\simeq s+\D_\phi+2 m\,, \qquad s\to \infty\,,
\label{deftwistlead}
\eeq
for asymptotically large $s$ and (non-negative) integer $m$. In fact, the entire $1/s$ expansion of the $\wh{\D}$ is computable once the bulk CFT data is known. In eq. \eqref{deftwistlead}, and in the rest of the paper, quantum numbers with a hat refer to the defect spectrum, except for the charge $s$.

It is not difficult to build some intuition for the large $s$ limit. Imagine to couple a $p$-dimensional CFT to a $d$-dimensional one via a weakly relevant operator for the lower dimensional theory. Further, assume that a short flow lands the system on a defect CFT. The $s=0$ sector of the defect spectrum is perturbatively close to the scaling dimensions of the original $p$-dimensional theory. On the other hand, operators which transform under transverse rotations are essentially bulk primaries evaluated at the location of the defect. The defect primaries in eq. \eqref{deftwistlead} are obtained by decomposing the conformal family of a scalar bulk primary in representations of $SO(p+1,1)\times SO(q-1,1)$, \ie, they are of the schematic form $\pa_i^s (\pa^j\pa_j)^m \phi$, where $i,\,j$ denote directions orthogonal to the defect. Therefore we shall call \emph{transverse derivative operators} the primaries whose scaling dimensions obey eq. \eqref{deftwistlead}. Operators of this kind also appear in the defect spectrum of large $N$ theories. The non-trivial statement, analogous to the case of double-twist operators in an ordinary CFT, is that their anomalous dimensions\footnote{We call ``anomalous dimension'' the deviation from eq. \eqref{deftwistlead}. This is perhaps non-standard, but hopefully not confusing.} are not only suppressed at weak coupling or large $N$, but also at large $s$. This expansion parameter is lost in the special case of a codimension one defect, \ie\, a boundary or an interface, transverse derivative operators still exist at small coupling or large $N$ in this case \cite{Rastelli:2017ecj}, however, we cannot constrain here their anomalous dimension in a generic CFT.\footnote{This does not exclude that a certain degree of universality appears in the large $\Dhat$ limit also for the spectrum of a boundary CFT \cite{Lauria:2017}.}

Equation \eqref{deftwistlead} can be obtained by lightcone bootstrap techniques, which we review and apply to this case in section \ref{sec:lightdefect}. In section \ref{sec:inversion}, we derive a Lorentzian inversion formula for the defect OPE \eqref{eq:inversionformula}, analogous to the one obtained in \cite{Caron-Huot:2017vep} for the four-point function of local operators -- see also the recent derivation in \cite{Simmons-Duffin:2017nub}.
The formula yields scaling dimensions of defect operators and bulk-to-defect OPE coefficients as analytic functions of the transverse spin, and thus implies the existence of trajectories in the $(\Dhat,s)$ space. It also resums the results that could be obtained by a systematic analysis of lightcone expansion, as done in \cite{Alday:2015ewa} for the case without defects.
However, \eqref{eq:inversionformula}, and thus the analytic constraint, is valid only for $s$ larger than a certain minimum $s_\star$. Unfortunately, contrary to \cite{Caron-Huot:2017vep}, we are not able to prove a theory independent upper bound to $s_\star$. We therefore ignore if there is a universal value of the transverse spin beyond which the spectrum of any defect CFT is constrained by analyticity. 

In section \ref{sec:ads}, we adopt a more physical point of view, and discuss the suppression of anomalous dimensions at large $s$ in a two-dimensional effective field theory, in a strict analogy with the work of Alday and Maldacena on large-spin operators \cite{Alday:2007mf}. This picture gives further intuition on additional accumulation points which are not of the form \eqref{deftwistlead}. In particular, we discuss adjoint insertions on Wilson lines, and we point out the connection between their scaling dimension and the inter-quark potential of a meson made of a heavy and a light (adjoint) quark. We predict the following asymptotic behavior:
\beq
\wh{\D} \simeq s+\frac{f(\lambda)}{2} \log s\,,
\label{singletracelead}
\eeq
up to $1/N$ and $1/s$ corrections. Here we used the standard $\mathcal{N}=4$ notation for the cusp anomalous dimension $f(\lambda)$, even if the result applies to a generic large-$N$ conformal gauge theory. The important point is that the coefficient of the logarithm is half of the one appearing in the twist of single-trace operators.

Section \ref{sec:examples} is devoted to the study of specific examples. We confirm the predictions of the lightcone bootstrap and of the inversion formula in various weakly-coupled scenarios, as well as in holography. 
In section \ref{sec:discussion} we summarize the main results and conclude with an outlook.

\section{Lightcone bootstrap with a defect}
\label{sec:lightdefect}

\begin{figure}[htb!]
\centering
\scalebox{0.5}{
\begin{pspicture}(0,-5.630351)(16.394688,5.630351)
\rput{-44.01924}(2.3641737,5.848707){\psframe[linewidth=0.04,linestyle=dashed,dash=0.16cm 0.16cm,dimen=outer](12.404317,3.9876986)(4.4289203,-3.9876986)}
\psdots[dotsize=0.4](13.981875,0.11978147)
\psdots[dotsize=0.4](10.941875,-2.1202185)
\psdots[dotsize=0.6,linecolor=red,dotstyle=otimes](2.781875,-0.10021853)
\rput{-45.0}(3.7411456,2.5483627){\rput(4.9485936,-3.2802186){\LARGE $z=0$}}
\rput{-45.0}(1.0685312,9.296725){\rput(11.758282,3.3197815){\LARGE $z=1$}}
\rput{45.0}(1.3022082,-9.456746){\rput(12.068281,-3.2002184){\LARGE $\bz=1$}}
\rput{45.0}(3.6794546,-2.515926){\rput(4.878594,3.1397815){\LARGE $\bz=0$}}
\rput(1.0385938,-0.36021852){\LARGE $z,\bz=0$}
\rput(15.41625,0.13978148){\LARGE $\phi(1,1)$}
\rput(10.06625,-1.5202185){\LARGE $\phi(z,\bz)$}
\rput(1.145,0.35978147){\LARGE defect}
\end{pspicture} 
}
\caption{The configuration of the insertions in $\braket{\phi(x_1)\phi(x_2)}$. We show a two-dimensional plane, transverse to the defect, where the two operators lie. The defect is space-like, it intersects the plane at the origin, and the two operators are placed at $(1,1)$ and $(z,\bz)$ respectively.}
\label{diamond}
\end{figure}
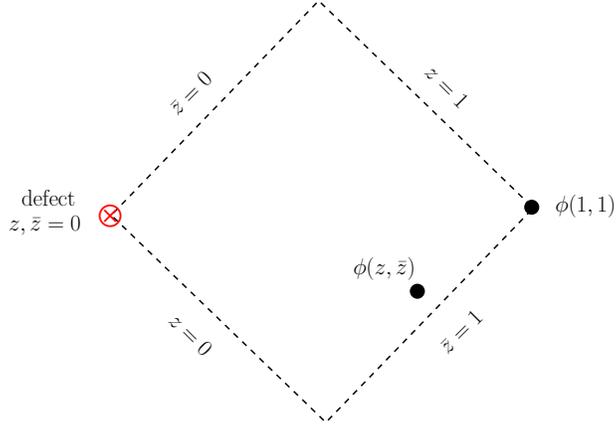

Let us recall the notation and  fix the setup. We consider a flat defect of codimension $q$ in $d$ spacetime dimensions. We shall also denote the dimension of the defect by $p$, \ie, $p+q=d$. The defect will always be space-like, except in section \ref{sec:ads}. Accordingly, we separate the spacetime indices ($\m=0,\dots, d-1$) in two subsets: orthogonal ($i=0,\dots, q-1$) and parallel ($a=q,\dots, d-1$) to the defect. 
Our main focus is the two-point function of identical scalar primaries which belong to the ambient CFT -- \emph{bulk primaries} for short. The correlator is a function of two cross-ratios, we refer to appendix \ref{sec:blocks} for some technical details and to ref.~\cite{Billo:2016cpy} for a general introduction to the topic of defect CFTs. Let us choose $x_{12}^a=0$, and focus on the two-dimensional plane in the transverse space which contains the origin and the two insertions. The geometry is shown in fig.~\ref{diamond}, and the cross-ratios can be traded for the lightcone coordinates $x_2=(z,\bz)$:
\beq
\braket{\phi(x_1)\phi(x_2)}=\frac{g(z,\bz)}{(\abs{x_1^i} \abs{x_2^i})^{\D_\phi}} =\frac{g(z,\bz)}{(z\bz)^{\D_\phi/2}} \,.
\label{2point}
\eeq
Some convenient features of the $(z,\bz)$ coordinates in the defect CFT context are discussed in \cite{Lauria:2017}.
The two-point function can be expanded either in the bulk or in the defect channel. The former involves the fusion of the two external operators, while the latter is obtained by separately fusing each of them with the defect. Agreement of the two expansions imposes the following crossing constraint:
\beq
g(z,\bz) =\left(\frac{(1-z)(1-\bz)}{(z\bz)^{1/2}}\right)^{-\D_\phi}\sum_{O} c_{\phi\phi O}\, a_O \,f_{\Delta,J}(z,\bz) =\sum_{\widehat{O}} (b_{\phi\widehat{O}})^2 \,\widehat{f}_{\widehat{\tau},s}(z,\bz)\,.
\label{eq:crossingsy}
\eeq
The first sum runs over the bulk spectrum, and $f_{\Delta,J}(z,\bz)$ is the bulk conformal block for the exchange of a primary of quantum numbers $(\Delta,J)$. The prefactor is chosen so that $f_{0,0}(z,\bz)=1$. The OPE data entering the bulk channel is the product of a three-point function coefficient ($c_{\phi\phi O}$) and the coefficient of the one-point function of the exchanged bulk operator ($a_O$). 
The second sum in eq.~\eqref{eq:crossingsy} runs over defect primaries. The latter do not carry $so(p)$ spin when the external operators are scalars, while a charge $s$ under the transverse $so(q-1,1)$ is allowed. The conformal blocks $\widehat{f}_{\widehat{\tau},s}(z,\bz)$ are then labeled by $s$ and $\widehat{\tau}=\widehat{\D}-s$ which we call the \emph{transverse twist}. 
The real numbers $b_{\phi\widehat{O}}$ determine the two-point function $\braket{\phi \widehat{O}}$.\footnote{We use a few slightly different conventions for the subscripts labeling $b_{\phi\widehat{O}}$ throughout the text depending on the context and expect that the notation is self-explanatory.}
The block in the defect channel is known exactly \cite{Billo:2016cpy} (see appendix \ref{sec:blocks} for the dictionary between our conventions and those of \cite{Billo:2016cpy}):
\beq
\fy_{\widehat{\tau},s}(z,\bz) = z^{\ttr/2}\bz^{\ttr/2+s}
{}_2F_1\left(-s,\frac{q}{2}-1,2-\frac{q}{2}-s,\frac{z}{\bz}\right) 
{}_2F_1\left(\ttr+s,\frac{p}{2},\ttr+s+1-\frac{p}{2},z\bz\right)\,.
\label{zdefblock}
\eeq
When $q$ is even, an order of limits ambiguity arises in the definition of the hypergeometric function, one must first take $s$ to be integer, and then $q$ to be even. This prescription is henceforth assumed. 
The bulk-channel conformal blocks are not known in closed-form for generic dimension and codimension. However, in $q=2$ and any $d$ \cite{Billo:2016cpy}, and $q=3$ and $d=4$ \cite{Liendo:2016ymz}, the blocks satisfy the same Casimir equation as the blocks of a four-point function of local operators, so any result in those cases carries over to the present situation.
In this work we are mostly interested in the lightcone limit (to be defined more precisely below), and in this limit the bulk blocks are given, for any $d$ and $q$, by\footnote{Notice that these are just the familiar (t-channel) $SL(2,\mathbb{R})$ blocks which appear in the lightcone decomposition of a four-point function of local operators. This is not hard to understand: the $SL(2,\mathbb{R})$ transformations act on the $z$ cross-ratios in the same way as they do in the case of a four-point function. They also act on the defect by simply displacing its intersection with the $z$ plane, as they would on a pair of local operators with $\D_1=\D_4=0$. Finally, the OPE limit itself is the usual lightcone OPE.
}
\beq
f_{\D,J}(z,\bz) = (1-\bz)^{\frac{\D-J}{2}} 
  \left( 2^{-J}\left(1-z\right)^{\frac{\D+J}{2}}
 {}_2 F_1 \left(\frac{\D+J}{2}, \frac{\D+J}{2}, \D+J, 1-z \right)+O((1-\bz))\right)\,.
 \label{singlecoll}
\eeq

One can ask if there can be a solution to the crossing equation \eqref{eq:crossingsy} with a finite number of blocks in either the bulk or defect decompositions. On the bulk side the answer is clearly yes: the trivial defect, \ie, the two-point function without a defect has a single bulk block, that of the identity, which is crossing symmetric. Whether it is possible to have a solution with finitely many non-trivial bulk primaries, on top of the identity, is a question that we do not address here.
On the defect side, it is not hard to prove that eq.~\eqref{eq:crossingsy} cannot be satisfied by finitely many defect primary operators if $q>1$. Indeed, the defect block in eq.~\eqref{zdefblock} has an unphysical singularity when $z\bz=1$, for any value of $\tfrac{z}{\bz}$, which is not consistent with the singularity structure of the two-point function in a Euclidean configuration. For $p >1 $ the behavior for $z \bz \to 1$ is
\beq
\begin{split}
&\fy_{\widehat{\tau},s}(z,\bz)  \overset{z\bz\to 1}{\sim}\\
& \frac{\Gamma\left(\frac{p-1}{2}\right) \Gamma\left(\widehat{\D}-\frac{p-2}{2}\right)}{2^{2-p}\sqrt{\pi} \Gamma(\widehat{\D})}
\left(\frac{\bz}{z}\right)^{s/2} 
{}_2F_1\left(-s,\frac{q}{2}-1,2-\frac{q}{2}-s,\frac{z}{\bz}\right) 
(1-z\bz)^{-p+1}\,,\quad p>1\,,
\end{split}
\label{sigmato1}
\eeq
while for $p=1$ the singularity is a logarithm, and the following argument is unchanged.
The case where $\Dhat = \tfrac{p-2}{2}$ looks different in \eqref{sigmato1}, but in this case too the argument goes through.
It can be easily checked that all the coefficients of the hypergeometric function in an expansion around $z=0$, which is a polynomial of degree $s$, are positive. Through the scalar unitarity bound, the sign of the prefactor in \eqref{sigmato1} is also fixed independently of the spectrum. Finally, positivity of the $(b_{\phi\widehat{O}})^2$ in eq.~\eqref{eq:crossingsy} implies that the singularity cannot be canceled by a finite number of blocks. In the case of a defect of codimension one the same singularity is instead potentially physical, and the exponent matches the exchange of the identity in the bulk channel when the theory is free. This allows for the existence of solutions to the boundary crossing equation with finitely many blocks \cite{McAvity:1995zd,Liendo:2012hy}.

It was shown in \cite{Pappadopulo:2012jk,Fitzpatrick:2012yx,Komargodski:2012ek} that analytic information can still be extracted from a crossing equation that contains infinitely many terms, by focusing on a limit that drastically simplifies one of the channels. Both in the Euclidean and in the lightcone OPE limits the identity dominates, say, the $t$-channel. This statement is theory independent, and is therefore the source of the universality of the spectrum needed in the $s$-channel to ensure crossing symmetry. In the case of the two-point function with a defect, there are two such simplifying limits. In the $z\to 0$ limit $\phi(x_2)$ is light-like separated from the defect, and the defect OPE is dominated by the operators with the smallest $\ttr=\wh{\D}-s$, as it can be seen in eq.~\eqref{zdefblock}. Conversely, when $(1-\bz) \to 0$ $\phi(x_2)$ is light-like separated from $\phi(x_1)$. This is the limit we will study in this section. When $(1-\bz) \to 0$ the identity in the bulk channel dominates the OPE, as it is clear from \eqref{singlecoll}, so one may hope that some analytic universal statement can be made about the defect spectrum. It remains to be proven that a specific sector of the defect spectrum is mostly sensitive to the identity in the crossed channel, so that a perturbation theory can be set up. We argue in this direction in subsection \ref{subsec:transder}, and also in subsection \ref{eq:CH_singlebulk} making use of the inversion formula. 

For the moment, we point out that such a sector of the defect spectrum needs to be perturbatively close to the spectrum of the trivial defect. Indeed, when there is no defect, translational invariance is preserved, and the identity is the only exchanged bulk primary
\beq
g(z,\bz)= \langle \phi(x_1) \phi(x_2) \rangle \left(z \bz\right)^{\Delta_\phi/2} = \left(\frac{\left(z \bz\right)^{1/2}}{(1-z)(1- \bz)}\right)^{\Delta_\phi}\,.
\label{eq:trivialdef}
\eeq

That the identity in the bulk channel is crossing symmetric on its own is a difference with respect to the four-point function: there the identity is never crossing symmetric. As anticipated in the introduction, the defect OPE of a real primary $\phi$ of dimension $\D_\phi$ is regular and only contains primaries of the kind
\beq
(\pa_i)^n(\pa_{j}\pa^j)^m \phi\,,\qquad 
\begin{cases}\wh{\D}=\D_\phi+n+2m\,, \\ s=n\,,
\end{cases}
\label{trivialspectrum}
\eeq
that is, derivatives of $\phi$ transverse to the defect. For later convenience, let us also report the defect OPE coefficients:
\beq
b^2_{s,m}=\frac{\Gamma \left(\frac{q}{2}+s\right) \Gamma (2 m+s+\D_\phi) \Gamma \left(m-\frac{d}{2}+\D_\phi+1\right)
   \Gamma \left(m-\frac{p}{2}+s+\D_\phi\right)}{\Gamma (\D_\phi) \Gamma (m+1) \Gamma (s+1) \Gamma
   \left(m+\frac{q}{2}+s\right) \Gamma \left(-\frac{d}{2}+\D_\phi+1\right) \Gamma \left(2 m-\frac{p}{2}+s+\D_\phi\right)}\,.
   \label{btrivial}
\eeq
We shall now argue that, at large $s$, 
the spectrum of any defect does contain a sector close to the trivial defect, in the same sense as ordinary CFT spectra are close to generalized free theory in the large spin limit. Our strategy in the rest of the section is analogous to the one presented in \cite{Komargodski:2012ek, Fitzpatrick:2012yx}. Now, certain weak points in the argument have been pointed out in \cite{Qiao:2017xif}. It may be expected that these issues are easier to tackle here with respect to the analogous ones in the lightcone bootstrap of the four-point function \cite{Komargodski:2012ek, Fitzpatrick:2012yx}, due to the simplicity of the defect channel blocks. However, in this work we do not try to solve them. While this section is not completely rigorous, it does define a calculable perturbative series, whose predictions we test in some examples in section \ref{sec:examples}. Furthermore, in subsection \ref{eq:CH_singlebulk} we shall come back to the question with a more rigorous tool in our hands, 
which will allow to rigorously prove part of the results that follow.


\subsection{The defect spectrum at large transverse spin}
\label{subsec:transder}

In what follows, we would like to analyze the crossing equation \eqref{eq:crossingsy} in the bulk light-cone limit, and more specifically in the following region:
\beq
\qquad 1-\bz\ll z < 1\,.
\label{looseregime}
\eeq

In this regime, the contribution of the higher-twist bulk operators is suppressed with respect to the identity \eqref{singlecoll}, while the defect OPE still converges. 
We can rewrite the crossing equation as follows:
\begin{equation}
1=\lim_{\bz\to1} \left(\frac{(1-z)(1-\bz)}{\sqrt{z\bz}}\right)^{\D_\phi}\sum_{\ttr,s} (b_{\phi\widehat{O}})^2 \,\widehat{f}_{\widehat{\tau},s}(z,\bz)\,.
\label{crossidentity}
\end{equation}
The conformal blocks of (a subsector of) the defect primaries need to match the $(1-\bz)$ dependence in the prefactor. However, each conformal block is analytic at $\bz=1$, as long as $z<1$, therefore $\lim_{\bz\to1}(1-\bz)^{\D_\phi}\widehat{f}_{\widehat{\tau},s}=0$ for every operator in the spectrum. We are led to conclude that the sum in eq.~\eqref{crossidentity} does not converge uniformly at $\bz=1$. Let us look for the region which is responsible for the singularity. At large and positive $\ttr$, for fixed $s$, the blocks are suppressed by $z^{\ttr/2}$ for every $z<1$ -- see eq.~\eqref{zdefblock}. Since $\ttr$ is not subject to a unitarity bound, one should worry about the $\ttr\to -\infty$ as well. In perturbation theory, we expect the transverse twist to be bounded from below. Furthermore, $\ttr$ generates time translations in AdS in the Alday-Maldacena picture explored in section \ref{sec:ads}, and this supports the more general assumption that $\ttr$  is bounded from below in a healthy theory. In the heuristic spirit of the discussion, we will not comment further on this limit.  
Finally, precisely in the limit $\bz\to1$ the sum over transverse spins, at fixed $\ttr$, ceases to be suppressed. In studying this last region, we replace the blocks with their large $s$ asymptotics:
\begin{equation}
\fy_{\widehat{\tau},s}(z,\bz) \underset{\ttr \textup{ fixed}}{\overset{s\to\infty}{\sim}}
(z\bz)^{\widehat{\tau}/2} \bz^s\left(\frac{\bz}{(\bz-z)}\right)^{\frac{q-2}{2}} 
(1-z\bz)^{-\frac{p}{2}}\left(1 + \OO\left(\frac{1}{s}\right)\right)\,.
\label{largesblocks}
\end{equation}
This approximation is obtained in the region $0<z<\bz<1$. Notice that the blocks are nicely factorized in the pairs $(z\bz,\ttr)$ and $(\bz,s)$ in this limit.  
If we now plug the asymptotics \eqref{largesblocks} in eq.~\eqref{crossidentity}, and we analyze the equation order by order in $z$, we see that eq.~\eqref{crossidentity} is satisfied if the following accumulation points exist in the spectrum:
\beq
\ttr = \D_\phi +2 m+ \OO(s^{-\alpha})~,\quad s\to \infty\,,
\label{ttrlead}
\eeq
for non-negative integer $m$, and a real positive $\alpha$ to be determined. The presence of a finite $s$ correction must be allowed, because we only treated the crossing equation \eqref{crossidentity} in the strict infinite $s$ limit. The operators \eqref{ttrlead} can obviously be thought of as the transverse derivatives described in eq.~\eqref{trivialspectrum}. Once we plug the spectrum \eqref{ttrlead} back in eq.~\eqref{crossidentity}, we also deduce that the OPE coefficients themselves should asymptote the ones in eq.~\eqref{btrivial}:
\beq
b^2_{s,m} =s^{\D_\phi-1} \left( \frac{1}{\Gamma(\D_\phi)} \binom{m-\frac{d}{2}+\D_\phi}{m} +\OO(s^{-\beta})\right)~,\quad s\to \infty~,
\label{btriviallead}
\eeq
for some positive $\beta$. 
  Let us emphasize what would be needed to make this argument rigorous -- see appendix F in \cite{Qiao:2017xif} for a more detailed discussion. One should prove that the following limit exists at fixed $\ttr$:
\beq
\rho(\ttr)=\lim_{\bz\to1} \sum_{s=0}^\infty (b_{\phi\widehat{O}})^2 \,(1-\bz)^{\D_\phi} \bz^s~,
\label{rhodensity}
\eeq
and that plugging $\rho(\ttr)$ in eq.~\eqref{crossidentity} one obtains a convergent sum over $\ttr$,\footnote{Up to the issue of bounding the spectrum at negative $\ttr$, this step can be done precisely as in \cite{Fitzpatrick:2012yx}.} so that eq.~\eqref{crossidentity}  can be analyzed order by order in $z$. The rest is then equivalent to the previous discussion: since the trivial defect in particular solves eq.~\eqref{crossidentity}, we obtain $\rho(\ttr)$ by plugging the OPE coefficient of the trivial defect in eq.~\eqref{rhodensity}. At this point, in turn, the Hardy-Littlewood tauberian theorem \cite{Qiao:2017xif} can be used to deduce from eq.~\eqref{rhodensity} the asymptotics \eqref{btriviallead}. Finally, we stress that eq.~\eqref{btriviallead} establishes an averaged property of the spectrum at large spin, while we have no control on the OPE coefficient of single defect primaries.\footnote{Let us also mention that the power-law form of the subleading corrections to eqs. (\ref{ttrlead}-\ref{btriviallead}) is assumed, but not implied even by the more rigorous approach. }

It is interesting to look in detail at the way the identity is reproduced at leading order in $1-\bz$. This highlights the relation between the large $s$ and small $1-\bz$ limits \cite{Komargodski:2012ek,Fitzpatrick:2012yx}. Let us write the r.h.s. of the crossing equation \eqref{eq:crossingsy} only including the transverse derivatives and using the large $s$ asymptotics of the blocks eq.~\eqref{largesblocks} and of the OPE coefficients eq.~\eqref{btriviallead}. We also replace the sum over spins by an integral, that we cutoff at some minimum spin $\Lambda$:
\begin{multline}
\left(\frac{\bz}{(\bz-z)}\right)^{\frac{q-2}{2}} 
(1-z\bz)^{-\frac{p}{2}} \sum_{m=0}^\infty \binom{m-\frac{d}{2}+\D_\phi}{m}  (z \bz)^{\D_\phi/2+m}
\left(\frac{1}{\Gamma(\D_\phi)}\int_\La^\infty ds s^{\D_\phi-1} \bz^s\right) \\
= \left(\frac{\sqrt{z\bz}}{1-z}\right)^{\D_\phi} \frac{\Gamma(\D_\phi,-\La \log \bz)}{\Gamma(\D_\phi)}\frac{1}{(-\log\bz)^{\D_\phi}}\,.
\label{largesdefchan}
\end{multline}
In the $\bz\to1$ limit, the result matches the bulk identity, for any finite $\La$, however large. This confirms that only asymptotically large values of $s$ matter. In fact, $\La$ could even be sent to infinity, as long as the growth is slower than $1/(1-\bz)$. This signals which range of spin is important in reproducing the bulk OPE limit, and cannot be excluded from the integral:
\beq
s \sim \frac{1}{1-\bz}\,.
\label{sbzrelation}
\eeq
An alternative way to understand this fact is through a saddle-point approximation of the simple integral in eq.~\eqref{largesdefchan}, as in \cite{Komargodski:2012ek}, which is accurate for large $\D_\phi$. The relation \eqref{sbzrelation} should be contrasted with the one relevant to double twist operators, that is $\ell\sim 1/(1-\bz)^{1/2}$ \cite{Komargodski:2012ek,Fitzpatrick:2012yx}. The different behavior here is responsible for the different finite spin exponent $\alpha$ -- see eq.~\eqref{ttrlead} below -- of the transverse derivatives with respect to the one of double twists.

Finite spin corrections to eqs. \eqref{ttrlead} and \eqref{btriviallead}  can be computed by taking into account subleading contributions to the bulk OPE in the $\bz\to 1$ limit.
A trivial series of corrections is required to match higher orders in $1-\bz$ coming from the bulk identity block. Only the OPE coefficients are affected, and all of the corrections are trivially obtained by expanding eq.~\eqref{btrivial} at large $s$. More interestingly, new bulk primaries start contributing at some order in $1-\bz$, according to their twist. For the sake of simplicity, we restrict the analysis to the correction due to a single bulk block. The case in which infinitely many primaries with (nearly) degenerate twist exist can also be dealt with, as done in \cite{Alday:2015ota,Alday:2016jfr} for the four-point function without defects, and is relevant to weakly coupled CFTs. We leave this analysis for future work. We assume that the leading contribution after the identity comes from a single bulk block with minimal twist $\tau_\textup{min}$ and dimension $\Delta_\textup{min}$:
\begin{equation}
\left(\frac{(z\bz)^{1/2}}{(1-z)(1-\bz)}\right)^{\D_\phi}
\bigr(1+c_{\phi\phi O_\textup{min}}\,a_\textup{min}\, f_{\D_\textup{min},J_\textup{min}}(z,\bz)\bigr) 
\sim 
\sum_{\substack{s\ \textup{large} \\ \wh{\D}-s\, =\, \ttr(s)}}\!\, b_{s,\ttr(s)}^2\,  \fy_{\ttr(s),s}(z,\bz)\,.
\label{limitcrossnext}
\end{equation}
We already assumed that in the regime $\eqref{looseregime}$ we can account for the leading twist operator $O_\textup{min}$ by modifying the trajectory $\ttr(s)$ of the transverse derivative operators and their OPE coefficient. Let us check that this is sufficient to reproduce the l.h.s., and let us only consider the leading transverse twist trajectory. This amounts to taking the small $z$ limit in the bulk collinear block \eqref{singlecoll}:
\beq
f_{\D,J}(z,\bz) \ 
\overset{z\to 0}{\sim}\ 
-2^{\D-1} \frac{\Ga\left(\frac{1}{2}+\frac{\D+J}{2}\right)}{\sqrt{\pi} \Ga\left(\frac{\D+J}{2}\right)} (1-\bz)^{\frac{\D-J}{2}}
 \left(2 \left(\gamma_E+\psi\left(\frac{\D+J}{2}\right)\right)+\log z\right)\,.
\label{doublecollall}
\eeq
Note that this contribution is only singular as $\bz\to1$ if $\Delta_\phi - \frac{\Delta-J}{2} <0$, in which case the rest of the discussion follows directly. We shall comment below on the opposite scenario.
Let us assume the following parametrization for the leading transverse twist trajectory:
\beq
\ttr(s)=\D_\phi+\frac{c_\textup{min}}{s^\al}\,.
\label{lightconetau}
\eeq
At large $s$, the anomalous dimension produces a logarithm of $z$ on the r.h.s. of eq.~\eqref{limitcrossnext}, which can be matched to the small $z$ behavior of the block eq.~\eqref{doublecollall}. A short computation yields:
\beq
\al =\frac{\tau_\textup{min}}{2} = \frac{\D_\textup{min}-J_\textup{min}}{2}\,,
\label{lightconeExponent}
\eeq
and
\beq
c_\textup{min}= - c_{\phi\phi O_\textup{min}}\,a_\textup{min} 2^{\D_\textup{min}}
 \frac{\Ga(\D_\phi)}{\Ga\left(\D_\phi-\frac{\tau_\textup{min}}{2}\right)}  \frac{\Ga\left(\frac{1}{2}+\frac{\D_\textup{min}+J_\textup{min}}{2}\right)}{\sqrt{\pi} \Ga\left(\frac{\D_\textup{min}+J_\textup{min}}{2}\right)}\,.
\label{cmin}
\eeq

Similarly, a correction to the OPE coefficient \eqref{btriviallead} is required to match the $\log$-independent part of eq.~\eqref{doublecollall}:
\begin{multline}
b_s^2=\frac{\Gamma(\Delta_\phi+s)}{\Gamma(s+1)\, \Gamma(\D_\phi)} 
\left(1+\frac{b_\textup{min}}{s^{\tau_\textup{min}/2}}\right), \\
b_\textup{min}= - c_{\phi\phi O_\textup{min}}\,a_\textup{min}\, 2^{\D_\textup{min}}
 \frac{\Ga(\D_\phi)}{\Ga\left(\D_\phi-\frac{\tau_\textup{min}}{2}\right)}  \frac{\Ga\left(\frac{1}{2}+\frac{\D_\textup{min}+J_\textup{min}}{2}\right)}{\sqrt{\pi} \Ga\left(\frac{\D_\textup{min}+J_\textup{min}}{2}\right)}\left(\gamma_E+\psi\left(\frac{\D_\textup{min}+J_\textup{min}}{2}\right)\right)\,.
 \label{bmin}
\end{multline}

Let us pause to comment on the non-singular case $\Delta_\phi - \frac{\Delta-J}{2} >0$. Following \cite{Alday:2015eya}, one can act with the defect Casimir operator $\Casdef$, written down in \cite{Billo:2016cpy}, on both sides of the crossing equation. On the bulk side we find
\beq
\Casdef \left[ (1-\bz)^\delta f(z)\right]= -2 \delta (\delta-1) (1-\bz)^{\delta-2}f(z) + \OO(1-\bz)^{\delta-1} \,,
\eeq
and so the leading behavior of \eqref{doublecollall} can be made singular by repeatedly acting with the defect Casimir, provided  $\tfrac{\Delta-J}{2}-\Delta_\phi$ is not a positive integer. For generic $\D$, $\D_\p$ the contribution of a bulk primary is thus Casimir-singular in the sense of \cite{Simmons-Duffin:2016wlq}. 
On the defect side, acting with $\Casdef$ introduces the eigenvalue for the corresponding defect block, which grows as $s^2$ for large $s$ and thus enhances the large $s$ behavior. Therefore, the results (\ref{cmin}-\ref{bmin}) are valid also if $\Delta_\phi - \frac{\Delta-J}{2}>0$ and non integer.

It is interesting to notice that unitarity does not fix the sign of $c_\textup{min}$. In other words, the spectrum does not need to be convex. However, it is intriguing to notice that in all the examples in section \ref{sec:examples} $c_\textup{min}<0$. In most of the cases the leading correction comes from the exchange of the stress-tensor block, which always turn out to have a negative coefficient of the one-point function: $a_T<0$. We shall comment more on this in subsection \ref{subsec:onepoint}.

Here we presented the result for the leading transverse twist correction, but similar corrections to \eqref{ttrlead} and \eqref{btriviallead} for $m \neq 0 $ are straightforward to obtain.
The large $s$ expansion of anomalous dimensions and OPE coefficients can be set up systematically to obtain the contribution of a collinear primary and all its descendants, as done in \cite{Alday:2015ewa}
for the four-point function case. The only requirement is the knowledge of the subleading contributions to \eqref{singlecoll}.
However, we shall pursue a different direction. In section \ref{sec:inversion}, we will obtain an inversion formula for the defect OPE, analogous to the one found in \cite{Caron-Huot:2017vep} for the four-point function, which allows to resum the lightcone expansion.

\section{Inversion of the defect OPE}
\label{sec:inversion}

In this section we describe a general way to extract the defect spectrum given a two-point function of bulk primaries. The quantum numbers $(\ttr,\,s)$ and the defect OPE coefficient $(b_{\phi\wh{O}})^2$ are extracted by an integral transform of the two-point function, which is analytic in the transverse spin $s$.
This is the defect analog of the inversion formula found in \cite{Caron-Huot:2017vep}, which applies to four-point functions in theories without defects, and most of the features of the present integral transform, and its derivation, are similar to \cite{Caron-Huot:2017vep}.
The inversion formula obtained in this section allows to resum the large $s$ results of section \ref{sec:lightdefect}, and extract the scaling dimension of defect operators with finite transverse spin. It also bypasses the need for some of the assumptions required by the lightcone analysis and discussed in section \ref{sec:lightdefect}.
The validity of the integral transform, similarly to that of \cite{Caron-Huot:2017vep}, depends on the growth of the correlator in a certain region. Contrary to \cite{Caron-Huot:2017vep}, though, the behavior of the correlator in this region is not controlled by an OPE limit, and we cannot place general bounds on its growth. We shall further comment on this issue in subsection \ref{sec:Lorentzianinv}.

In the rest of this section we derive the inversion formula for the defect OPE following in the footsteps of \cite{Caron-Huot:2017vep}. For this purpose we start by obtaining a Euclidean inversion formula, which simply follows from orthogonality of partial waves, see for example \cite{Hogervorst:2017sfd} for a detailed derivation in the case of a four-point function in one-dimensional theories, or \cite{Hogervorst:2017kbj} for boundary CFTs. While this Euclidean formula is not analytic in the transverse spin $s$, it can be manipulated into a Lorentzian formula that is.
A different derivation of the Lorentzian inversion formula without defects was presented recently in \cite{Simmons-Duffin:2017nub}: we leave to future work the extension of that physically more transparent approach to the present case.


\subsection{The Euclidean formula}

We start by obtaining an Euclidean inversion formula for the defect OPE. 
Recall that in our configuration the two operators lie on a plane orthogonal to the defect. The defect intersects the plane at the origin, with one external operator placed at $x_1=(1,1)$ and the other at $x_2=(z,\bz)$, see fig.~\ref{diamond}. 
We introduce the following radial coordinates for the position of the second operator
\beq
z=r w\,,\qquad \bz=\frac{r}{w}\,,\qquad \eta= \frac{1}{2}\left(w + \frac{1}{w}\right)\,,
\label{utoradial}
\eeq
where in Euclidean signature $\bz=z^*$ and so $w$ is a phase. 
Since defect blocks \eqref{zdefblock} factorize,
\begin{align}
\fy_{\widehat{\tau},s}(z,\bz) =\wh{g}_s(\eta)\fy_\Dhat (r)\,, 
\label{eq:factorizedblock}
\end{align}
where $\widehat{\tau}=\Dhat-s$, we shall treat the parallel $\fy_\Dhat(r)$ and angular $\wh{g}_s(\eta)$ parts separately.

\subsubsection*{Parallel factor of defect blocks}

Let us start by considering the parallel factor of the block. Similarly to \cite{Hogervorst:2017sfd}, we re-write the Casimir equation that $\fy_\Dhat(r)$ satisfies \cite{Billo:2016cpy} in the form of a Sturm-Liouville problem
\beq
\SLDr \fy(r) =  \Dhat (\Dhat-p)\fy(r)\,, \qquad \mathrm{with}\qquad \SLDr \fy(r)= \frac{r^{p+1}}{(1-r^2)^p} \frac{d}{dr}\left(r^{1-p} (1-r^2)^p \;  \frac{d\fy(r)}{dr}\right)\,.
\label{eq:Casturm}
\eeq
The operator $\SLDr$ defined in \eqref{eq:Casturm} is self-adjoint with respect to the measure
\beq
\mu_p(r)=\frac{(1-r^2)^p}{r^{p+1}}\,,
\label{mupar}
\eeq
in the interval $r\in [0,1]$,
provided the functions are well-behaved near $r=0$ and $r=1$. Concretely, self-adjointness requires that the following boundary term vanishes
\beq
\begin{split}
&\int\limits_0^1 dr\, \m_p(r)\,\SLDr(\Psi(r))\tilde{\Psi}(r)-\int\limits_0^1 dr \, \m_p(r)\,\Psi(r)\SLDr(\tilde{\Psi}(r)) \\
&\qquad\qquad\qquad= \int\limits_0^1 dr\,\frac{d}{dr}\left[\m_p(r) r^2\left( \frac{d\Psi(r)}{dr}  \tilde{\Psi}(r)-\Psi(r)   \frac{d\tilde{\Psi}(r)}{dr} \right)\right]\,.
\end{split}
\label{eq:selfadj}
\eeq
While for the functions to be square-integrable with respect to the measure \eqref{mupar} their behavior near $r=0$ and $r=1$ must be such that
\beq
\Psi(r) \underset{r \to 0}{\sim} r^{\frac{p}{2}+\epsilon} \,, \qquad \Psi(r) \underset{r \to 1}{\sim} (1-r)^{-\frac{p+1}{2}+\epsilon^\prime}\,,
\eeq
with $\epsilon$, $\epsilon^\prime$ positive numbers.
However, the parallel factor in the defect conformal blocks,
\beq
\fy_\Dhat(r) = r^{\Dhat}\; {}_2F_1\left(\Dhat,\frac{p}{2},\Dhat+1-\frac{p}{2}, r^2 \right)\,,
\eeq
which is an eigenfunction of $\SLDr$,
grows as $(1-r)^{1-p}$ for $r \to 1$ (this growth is logarithmic in the $p=1$ case). Therefore, unless $p=1,2$ their square is not integrable against the measure \eqref{mupar}, and for no value of $p$ does the boundary term in \eqref{eq:selfadj} vanish. 

Following \cite{Hogervorst:2017sfd,Caron-Huot:2017vep} we consider a linear combination of $\fy_\Dhat$ that is still an eigenfunction of $\SLDr$, with eigenvalue $\Dhat(\Dhat-p)$, but is regular at $r=1$
\beq
\Psi_\Dhat(r)=\frac{1}{2}\left(\fy_{\Dhat}(r)+\frac{K_{p-\Dhat}}{K_\Dhat}\fy_{p-\Dhat}(r)\right)= \frac{K_{p-\Dhat}}{2 K_p} r^{p-\Dhat}\;
{}_2F_1 \left(\frac{p}{2},p-\Dhat ,p,1-r^2\right)\,,
\label{bdelta}
\eeq
where we defined
\beq
K_\Dhat=\frac{\Ga(\Dhat)}{\Gamma\left(\Dhat-\frac{p}{2}\right)}\,.
\eeq
Also, the behavior of $\Psi_\Dhat(r)$ near $r=1$ is such that the corresponding boundary term in \eqref{eq:selfadj} vanishes.
However, near $r=0$ the functions grow as 
\beq
\Psi_\Dhat(r) \underset{r \to 0}{\sim} \frac{\Gamma (p-\Dhat) \Gamma \left(\Dhat-\frac{p}{2}\right) r^{p-\Dhat}}{2 \Gamma (\Dhat) \Gamma \left(\frac{p}{2}-\Dhat\right)}+\frac{r^{\Dhat}}{2}\,,
\label{eq:bdeltanear0}
\eeq
and so at best they can be delta-function normalizable, provided we take $\Re(\Dhat)=\tfrac{p}{2}$. Were we to consider normalizable eigenfunctions of the self-adjoint operator $\SLDr$, standard arguments would imply that they are orthogonal. To show orthogonality in this case, we will work instead with the following regularized functions
\beq
\Psi_\Dhat^{\mathrm{reg.}}(r) = \frac{K_{p-\Dhat}}{2 K_p} r^{p-\Dhat+\epsilon}\; {}_2F_1 \left(\frac{p}{2},p-\Dhat ,p,1-r^2\right)\,,
\qquad \mathrm{with}\;\;
\Dhat = \frac{p}{2} + \ii \, \nu\,, \quad \nu \in  \mathbb{R}\,,
\label{eq:bdeltareg}
\eeq
with $\epsilon >0$ a small number, such that the functions $\Psi_\Dhat^{\mathrm{reg.}}(r)$ are normalizable. The operator  $\SLDr$ is self-adjoint on these functions, since the chosen regularization makes the boundary term at $r=0$ vanish, while it preserves the vanishing of the boundary term at $r=1$.
Due to the regularization $\Psi_\Dhat^{\mathrm{reg.}}(r)$ are not eigenfunctions of $\SLDr$ and so orthogonality is not yet immediate. Nevertheless, starting from self-adjointness (the first line in \eqref{eq:orthogonalityproof}), we can evaluate the action of $\SLDr$ on the regularized functions to obtain 
\beq
\begin{split}
0  &= \int\limits_0^1 dr\, \m_p(r)\,\left(\SLDr(\Psi^{\mathrm{reg.}}_{\Dhat_1}(r))\Psi^{\mathrm{reg.}}_{\Dhat_2}(r)-\Psi^{\mathrm{reg.}}_{\Dhat_1}(r) \SLDr(\Psi^{\mathrm{reg.}}_{\Dhat_2}(r))\right) \\
 &= \left(\Dhat_1 (\Dhat_1-p)-\Dhat_2 (\Dhat_2-p)\right)\; \int\limits_0^1 dr\, \m_p(r)\,\Psi^{\mathrm{reg.}}_{\Dhat_1}(r) \Psi^{\mathrm{reg.}}_{\Dhat_2}(r) + \OO(\epsilon)\,.
\end{split}
\label{eq:orthogonalityproof}
\eeq
Taking $\epsilon \to 0 $ this implies that if $\Dhat_1 (\Dhat_1-p) \neq \Dhat_2 (\Dhat_2-p)$ the functions are orthogonal.
Finally, all we have to show now is what happens when the eigenvalues coincide, and for that we need only examine the behavior of the functions near $r=0$ where the integral develops a singularity. In this case, taking $\Dhat_i=p/2+\ii \, \n_i$, we end up with integrals of the type 
\beq
\int\limits_0 dr\, r^{-1 \pm \ii (\nu_1 \pm \nu_2)} = \pi \delta(\nu_1 \pm \nu_2) + \text{non-singular}\,,
\eeq
following from the behavior of the measure \eqref{mupar} and  \eqref{eq:bdeltanear0}.\footnote{One could equivalently have shown that the integral of the regularized functions \eqref{eq:bdeltareg}  provides a representation of the delta function as $\epsilon\to0$, this is obvious for $p=2$ when the resulting expressions are very simple.}

All in all, the functions \eqref{bdelta} are orthogonal when $\Dhat_i=p/2+\ii \, \n_i$, satisfying\footnote{The functions $\Psi_\Dhat(r)$ could be made real for $\n \in \mathbb{R}$ by an appropriate choice of normalization, but we have not done so. Also, the orthogonality of \eqref{orthopar} is enough for our purposes and thus we do not define a positive inner product.}
\beq
\int\limits_0^1dr \, \m_p(r)\,\Psi_{\Dhat_1}(r)\Psi_{\Dhat_2}(r) = \frac{\pi}{2}\frac{K_{p-\Dhat_2}}{K_{\Dhat_1}}\left[\delta(\n_1-\n_2)+\delta(\n_1+\n_2)\right]\,.
\label{orthopar}
\eeq

\subsubsection*{Angular factor of defect blocks}

We now turn to the angular factor in the conformal block \eqref{eq:factorizedblock}. It is useful to go back to the representation of the angular factor in \eqref{zdefblock} as a Gegenbauer polynomial for integer $s$ via
\beq
w^{-s} {}_2F_1\left(-s,\frac{q}{2}-1,2-\frac{q}{2}-s,w^2\right)=\binom{s+\frac{q}{2}-2}{\frac{q}{2}-2}^{-1} C_s^{q/2-1}\left(\frac{w}{2}+\frac{1}{2 w}\right)\,,
\label{gegenhyper}
\eeq
such that it becomes
\beq
\wh{g}_s(\eta) = \binom{s+\frac{q}{2}-2}{\frac{q}{2}-2}^{-1}\, C_s^{(q/2-1)}(\eta) \,. 
\label{eq:geta}
\eeq
Gegenbauer polynomials are orthogonal with respect to the following measure
\beq
\int\limits_{-1}^1d\eta\, \mu_q(\eta) C_s^{(\frac{q}{2}-1)}(\eta) C_{s'}^{(\frac{q}{2}-1)}(\eta)
= \frac{ 2^{3-q}\pi\Gamma(s+q-2)}{\left(s+\frac{q}{2}-1\right)\Gamma(s+1) \Gamma\left(\frac{q}{2}-1\right)^2}\de_{ss'}\,, \quad
\mu_q(\eta)=(1-\eta^2)^\frac{q-3}{2}\,,
\label{orthogeg}
\eeq
which we rewrite using the normalization of the conformal block themselves
\beq
\int\limits_{-1}^1d\eta\, \mu_q(\eta) \wh{g}_s(\eta) \wh{g}_{s'}(\eta)
= N_{q,s}\de_{ss'}\,,\qquad 
N_{q,s}=2^{3-q}\pi\frac{\Gamma(s+1) \Gamma(s+q-2)}{\Gamma(s+\frac{q}{2}) \Gamma\left(s+\frac{q}{2}-1\right)}\,.
\label{orthoangblocks}
\eeq

\subsubsection*{Euclidean inversion formula}
Finally we can write the following orthogonal decomposition for the two-point function, similarly to what has been done for the case of the four-point function of local operators \cite{Dobrev:1975ru} (see also \cite{Costa:2012cb}),
\beq
g(r,\eta)=\sum\limits_{s=0}^\infty \int_\gamma \frac{d\Dhat}{2\pi\ii} b(\Dhat,s)\wh{g}_s(\eta) \Psi_\Dhat(r),\qquad \ga=\{\Dhat: \Dhat \in \left( p/2 - \ii \infty, p/2 + \ii \infty \right) \, \}\,.
\label{2pointdec}
\eeq
Since $\Psi_{p-\Dhat}(r)=\frac{K_\Dhat}{K_{p-\Dhat}} \Psi_\Dhat(r)$, we can assume that
\beq
b(p-\Dhat,s)= \frac{K_{p-\Dhat}}{K_\Dhat} b(\Dhat,s)\,.
\label{eq:bmirror}
\eeq
The position of the poles and residues of $b(\Dhat,s)$ is revealed by closing the contour $\gamma$. 
At large~$\Dhat$
\beq
\fy_\Dhat(r)\sim r^\Dhat(1-r^2)^{-p/2}\,,
\eeq
so the contour must be closed to the right on the first addend in $\Psi_\Dhat$, and to the left on the second - see eq.~\eqref{bdelta}. In order for the result to agree with the usual conformal block decomposition, $b(\Dhat,s)$ must have single poles in correspondence of the spectrum, and the residue must coincide, up to a sign, with the OPE coefficient:\footnote{For defect operators of dimension less than $p/2$ we  must deform the contour such that it picks up the pole on the left and does not pick up the reflection according to \eqref{eq:bmirror} on the right. Similarly if the operator has dimension exactly $p/2$ we must take the principle-value of the integral to pick up  half of the residue.}
\beq
g(r,\eta)=\sum\limits_{s=0}^\infty \wh{g}_s(\eta) \sum_{\Dhat^*\in \textup{spectrum}}
b_{s,\Dhat^*}^2 \fy_{\Dhat^*}(r)\,, \qquad b_{s,\Dhat^*}^2=-\textup{Res}_{\Dhat=\Dhat^*} b(\Dhat,s)\,.
\label{eq:resb}
\eeq
Not all poles in \eqref{2pointdec} arise from  poles of $b(\Dhat,s)$, as the defect blocks themselves have poles for special values of $\Dhat$ and  $s$.
However, since the defect blocks $\fy_{\Dhat}(r)$ have poles for $\Dhat= \tfrac{p}{2}-n$ \cite{Lauria:2017} they are always to the left of $\tfrac{p}{2}$, and thus are not picked up when we close the contour to the right. Similarly, for the second addend in $\Psi_\Dhat$, we close the contour to the left while $\fy_{p-\Dhat}(r)$ only has poles to the right of $\tfrac{p}{2}$.

Eq.~\eqref{2pointdec} can be easily inverted using the orthogonality relations \eqref{orthopar} and \eqref{orthoangblocks}, yielding the following Euclidean inversion formula\footnote{We thank D.~Maz\'{a}\v{c} for collaboration in obtaining this formula.}
\beq
b(\Dhat,s)=\frac{2}{N_{q,s}} \frac{K_\Dhat}{K_{p-\Dhat}}\;\int\limits_{-1}^{1}d\eta
\int\limits_{0}^1 dr\, \mu_p(r) \mu_q(\eta) \, \wh{g}_s(\eta)\Psi_\Dhat(r) g(r,\eta)\,.
\eeq
Since in Euclidean signature $\eta$ is nothing more than the cosine of an angle (see \eqref{utoradial}), we change variables in the above integral to obtain
\beq
\begin{split}
b(\Dhat,s)&=\frac{1}{N_{q,s}} \frac{K_\Dhat}{K_{p-\Dhat}}\oint\limits_{|w|=1} \frac{dw}{\ii w}
\int\limits_{0}^1 dr\, \mu(r,w) \, \wh{g}_s\left(\frac{1}{2w}+\frac{w}{2}\right)\Psi_\Dhat(r) \; g\left(r,\frac{1}{2w}+\frac{w}{2}\right)\,, \\
\mu(r,w)&=\mu_p(r) \left| \frac{w}{2 \ii} - \frac{1}{2 \ii w}\right|^{q-2}\,.
\end{split}
\label{euclideaninv}
\eeq
The above integral might not converge. Let us first consider the region $r\sim 0$. Here, convergence is controlled by the scaling dimension $\Dhat_\textup{min}$ of the lightest defect primary exchanged in $g(r,\eta)$. Specifically, the integral converges in the strip
\beq
\left|\Re\Dhat-\frac{p}{2}\right|<\Dhat_\textup{min}-\frac{p}{2}~.
\label{Euclidstrip}
\eeq
To obtain an analytic function in $\Dhat$, it is therefore necessary to subtract from $g(r,\eta)$ the blocks\footnote{Actually, in order to preserve the analytic structure of the two-point function at $r=1$ for generic $\eta$, one should subtract the corresponding $\Psi$ rather than the block. See appendix B.2 of \cite{Simmons-Duffin:2017nub} for more details.} corresponding to operators lighter than $p/2$, and invert the subtracted two-point function. If necessary, the poles corresponding to these operators can be added by hand to $b(\Dhat,s)$. Once the strip \eqref{Euclidstrip} exists, the function $b(\Dhat,s)$ can be analytically continued past its edges as follows. Suppose that we want to extend towards the right -- going towards the left is similar. For $\Dhat>p/2$, the kernel of eq. \eqref{euclideaninv} has an expansion in growing powers of $r$ that starts with $r^{-1-\Dhat}$. Then we subtract the leading term $g(r,\eta)\sim r^{\Dhat_\textup{min}}$, and integrate it separately, defining the result by analytic continuation in $\Dhat$:
\beq
\int_0^1\!dr\, r^{-1-\Dhat+\Dhat_\textup{min}}=\frac{1}{\Dhat_\textup{min}-\Dhat}~.
\label{EucliDeltapoles}
\eeq
This defines the sought analytic continuation of $b(\Dhat,s)$ up to the next exchanged operator. Proceeding order by order in $r$, one constructs the full function $b(\Dhat,s)$.\footnote{In fact, at $\Dhat=p/2+n$, for integer $n>0$ the combination $K_\Dhat/K_{p-\Dhat}\Psi_\Dhat$, which appears in the inversion formula \eqref{euclideaninv}, develops a $\log r$ at small $r$. This is due to the cancellation of a pole in the difference between the block and its shadow. This is harmless in general, but extra care is needed if an operator with dimension $p/2+n$ exists in the spectrum.}

Divergences at $r=1$, on the other hand, are controlled by the bulk channel OPE -- in particular, they are absent if $\D_\phi<p/2+1$. They may be regulated by cutting off the integral at $r<1-\epsilon$. The kernel $\mu(r,w) \Psi_\Dhat(r)$ has a regular Taylor expansion close to $r=1$, therefore the divergent part of the inversion formula as $\epsilon\to0$ does not contain poles in $\Dhat$. Hence, it can be safely dropped without altering the spectrum and the residues of $b(\Dhat,s)$.


\subsection{The Lorentzian formula}
\label{sec:Lorentzianinv}
In the Euclidean inversion formula \eqref{euclideaninv} the contour of integration in the complex $w$ plane is the unit circle, as $w$ is a phase. We now want to deform the contour in order to integrate over real values of $w$, which correspond to a Lorentzian configuration.
The range of $r$ in the Euclidean formula is confined between $0$ and $1$.
\begin{figure}[htb!]
\centering
\scalebox{0.6} 
{
\begin{pspicture}(0,-5.6306376)(12.46,5.6306376)
\psdots[dotsize=0.3](9.56,0.04936256)
\psdots[dotsize=0.6,linecolor=red,dotstyle=otimes](3.94,-0.11063744)
\rput{-45.0}(4.1262264,3.66813){\rput(6.4709377,-3.1506374){\LARGE $z=0$}}
\rput{45.0}(4.390353,-3.7824004){\rput(6.7409377,3.4093626){\LARGE $\bar{z}=0$}}
\rput(2.2609375,-0.35063744){\LARGE $z,\bar{z}=0$}
\rput(2.3609376,0.36936256){\LARGE $\text{defect}$}
\psline[linewidth=0.04cm,linestyle=dashed,dash=0.16cm 0.16cm](3.9613216,-0.111912325)(9.478679,5.6106377)
\psline[linewidth=0.04cm,linestyle=dashed,dash=0.16cm 0.16cm](6.7413216,-2.8519123)(12.258678,2.8706374)
\psline[linewidth=0.04cm,linestyle=dashed,dash=0.16cm 0.16cm](6.74,2.7693624)(12.44,-2.7506375)
\psline[linewidth=0.04cm,linestyle=dashed,dash=0.16cm 0.16cm](3.96,-0.09063744)(9.66,-5.6106377)
\rput{90.16892}(9.904397,-9.743365){\psarc[linewidth=0.04](9.80954,0.06593688){3.193062}{54.998444}{125.7219}}
\psline[linewidth=0.04cm](7.18,1.8893626)(8.8,4.3493624)
\psline[linewidth=0.04cm](7.2103586,-1.7848043)(8.929642,-4.1764703)
\rput(9.774531,4.7){\LARGE $w \to \infty$}
\rput(9.784532,-4.6606374){\LARGE $w \to 0$}
\rput(8.884531,-2.1606374){\LARGE $w=r$}
\rput(8.764531,2.1593626){\LARGE $w=\frac{1}{r}$}
\end{pspicture} 
}
\caption{The positive real axis on the $w$ complex plane, at fixed $r<1$, maps to the black solid line in the $(z,\bz)$ plane. The bulk OPE singularities correspond to the intersection of the line with the past and future lightcones of the operator $\phi(1,1)$.}
\label{bla}
\end{figure}
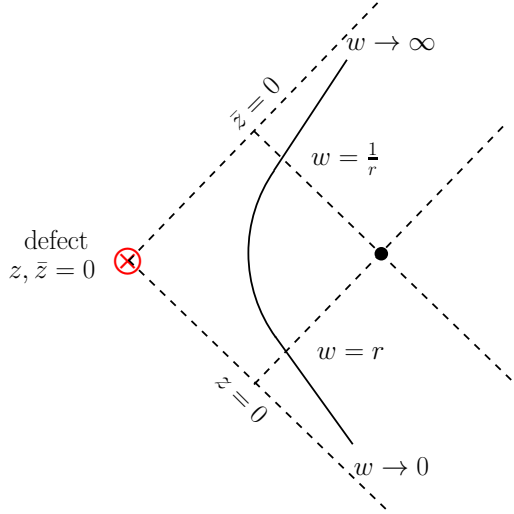
All other points in the Euclidean plane are related to this fundamental region by inversion. As it is clear from figure \ref{bla}, at fixed $r<1$ the function $g(r,w)$ has two copies of the bulk OPE singularity at $w=r$ and $w=1/r$. Contrary to the case of the four-point function, there is no singularity at negative values of $w$. Further singularities may lie in the limits $w=0$ and $w=\infty$, which are double lightcone limits -- see the explanation in appendix \ref{sec:rhobulk} and fig.~\ref{fig:rhobulk} there. The OPE singularities are generically branch points, while we know that the correlator is single valued on the circle $|w|=1$. Therefore the cuts run from 0 to $r$ and from $1/r$ to $\infty$. The remaining $w$ dependence in the inversion formula comes from $\wh{g}_s$ and the measure.
It is useful to consider a larger set of solutions to the Casimir equation associated to the orthogonal rotations. 
We make use of \eqref{gegenhyper} again to go back to a representation of $\wh{g}_s(\eta)$ \eqref{eq:geta} as a hypergeometric function
\beq
\wh{h}_1(s,w)= \binom{s+\frac{q}{2}-2}{\frac{q}{2}-2}^{-1} C_s^{q/2-1}\left(\frac{w}{2}+\frac{1}{2 w}\right)=
w^{-s} {}_2F_1\left(-s,\frac{q}{2}-1,2-\frac{q}{2}-s,w^2\right)\,.
\label{eq:h1def}
\eeq
Recall that when $q$ is even, an order of limits ambiguity arises in the definition of the hypergeometric function. The equality \eqref{gegenhyper} holds if we first take $s$ to be integer, and then $q$ to be even. As before, this prescription is assumed every time it is necessary.

Other solutions to the Casimir equation for the angular part of the defect blocks can be obtained by combining the transformations $w\to1/w$ -- which leaves $\wh{h}_1$ invariant when $s$ is integer -- and $s\to 2-q-s$, both of which are symmetries of the Casimir equation. We will use the two following solutions
\begin{align}
\wh{h}_2(s,w) &\colonequals \wh{h}_1(2-q-s,w)=
w^{s+q-2} {}_2F_1\left(s+q-2,\frac{q}{2}-1,\frac{q}{2}+s,w^2\right)\,,\\
 \wh{h}_3(s,w) &\colonequals \wh{h}_2(s,1/w) =
 w^{2-q-s} {}_2F_1\left(s+q-2,\frac{q}{2}-1,\frac{q}{2}+s,\frac{1}{w^2}\right)\,,
\end{align}
where $\wh{h}_2$ is regular at the origin while $\wh{h}_3$ is regular at infinity. Ideally, one would like to express $\wh{h}_1$ as a linear combination of $\wh{h}_2$ and $\wh{h}_3$, but this is globally possible only for defects of even codimension. Indeed, when $q$ is even, the discontinuities of $\wh{h}_2$ and $\wh{h}_3$ vanish. Let us first consider this simpler case.

\subsubsection*{\texorpdfstring{Even $q$}{Even q}}

\begin{figure}[tb!]
\centering
\scalebox{0.6} 
{
\begin{pspicture}(0,-4.66)(20.6,4.66)
\psline[linewidth=0.04cm,arrowsize=0.05291667cm 2.0,arrowlength=1.4,arrowinset=0.4]{<-}(8.96,0.16)(0.0,0.16)
\psline[linewidth=0.04cm,arrowsize=0.05291667cm 2.0,arrowlength=1.4,arrowinset=0.4]{<-}(4.48,4.64)(4.48,-4.32)
\pscircle[linewidth=0.022,linestyle=dashed,dash=0.16cm 0.16cm,dimen=outer](4.48,0.16){1.41}
\psline[linewidth=0.04](8.26,4.36)(8.26,4.04)(8.26,4.04)(8.62,4.04)(8.58,4.04)
\rput(8.5,4.23){\LARGE $w$}
\rput(6,0.51){\LARGE $1$}
\rput(2.75,0.49){\LARGE $-1$}
\rput(7.42,-0.36){\LARGE $\tfrac{1}{r}$}
\psline[linewidth=0.02cm](1.54,0.24)(1.54,0.06)
\psline[linewidth=0.02cm](7.42,0.24)(7.42,0.06)
\psarc[linewidth=0.03](4.49,0.29){3.73}{0.0}{180.0}
\psline[linewidth=0.03cm](0.75,0.3)(8.23,0.3)
\psline[linewidth=0.02](6.018337,3.859643)(5.634211,3.8303933)(5.927452,3.5528207)
\rput{-180.0}(8.98,0.06){\psarc[linewidth=0.03](4.49,0.03){3.73}{0.0}{180.0}}
\psline[linewidth=0.03cm](0.75,0.02)(8.23,0.02)
\psline[linewidth=0.02cm](5.17,0.24)(5.17,0.06)
\psline[linewidth=0.02cm](3.79,0.26)(3.79,0.08)
\rput(5.1753125,-0.21){\LARGE $r$}
\psline[linewidth=0.04cm,arrowsize=0.05291667cm 2.0,arrowlength=1.4,arrowinset=0.4]{<-}(20.58,-0.16)(11.62,-0.16)
\psline[linewidth=0.04cm,arrowsize=0.05291667cm 2.0,arrowlength=1.4,arrowinset=0.4]{<-}(16.1,4.32)(16.1,-4.64)
\psline[linewidth=0.04](19.88,4.04)(19.88,3.72)(19.88,3.72)(20.24,3.72)(20.2,3.72)
\rput(20.15,3.91){\LARGE $w$}
\rput(17.858126,-0.57){\LARGE $1$}
\rput(19.135157,-0.82){\LARGE $\tfrac{1}{r}$}
\psline[linewidth=0.02cm](19.04,-0.08)(19.04,-0.26)
\psline[linewidth=0.02cm](16.85,-0.06)(16.85,-0.24)
\psline[linewidth=0.02cm](17.89,-0.04)(17.89,-0.22)
\rput(16.85,-0.58){\LARGE $r$}
\rput(20.15,3.91){\LARGE $w$}
\rput(17.858126,-0.57){\LARGE $1$}
\rput(19.135157,-0.82){\LARGE $\tfrac{1}{r}$}
\psline[linewidth=0.04cm,arrowsize=0.05291667cm 3.0,arrowlength=1.4,arrowinset=0.4]{->}(9.42,1.66)(11.04,1.66)
\psline[linewidth=0.04](16.1,-0.14)(16.16,0.0)(16.24,-0.26)(16.32,0.0)(16.4,-0.26)(16.48,0.0)(16.54,-0.26)(16.62,0.0)(16.7,-0.26)(16.78,0.0)(16.82,-0.14)(16.82,-0.14)
\psline[linewidth=0.04](19.04,-0.18)(19.1,-0.04)(19.18,-0.3)(19.26,-0.04)(19.34,-0.3)(19.42,-0.04)(19.48,-0.3)(19.56,-0.04)(19.64,-0.3)(19.72,-0.04)(19.8,-0.3)(19.88,-0.04)
\psellipse[linewidth=0.04,dimen=outer](16.47,-0.11)(0.57,0.39)
\psarc[linewidth=0.04](19.1,-0.16){0.28}{90.0}{270.0}
\psline[linewidth=0.04cm]{cc-}(19.04,0.12)(20.48,0.12)
\psline[linewidth=0.04cm](19.08,-0.44)(20.48,-0.44)
\psline[linewidth=0.04](19.5,-0.16)(19.56,-0.02)(19.64,-0.28)(19.72,-0.02)(19.8,-0.28)(19.88,-0.02)(19.94,-0.28)(20.02,-0.02)(20.1,-0.28)(20.18,-0.02)(20.26,-0.28)(20.34,-0.02)
\rput(5.5915623,3.33){\LARGE $C_+$}
\rput(5.7159376,-2.87){\LARGE $C_-$}
\psline[linewidth=0.02](6.5260234,-3.286245)(6.7136455,-2.9497848)(6.320331,-3.0411117)
\psline[linewidth=0.02](5.2193084,1.5382569)(4.834942,1.5123606)(5.1257496,1.2322396)
\psline[linewidth=0.02](16.78,0.4)(16.42,0.26285714)(16.78,0.08)
\psline[linewidth=0.02](19.14,-0.024761964)(19.5,0.11238095)(19.14,0.2952381)
\end{pspicture} 
}

\caption{Deformation of the $|w|=1$ contour of \eqref{euclideaninv} used to define $b_0(\Dhat,s)$ and $b_\infty(\Dhat,s)$ in \eqref{lorinveven}. For the case of odd codimension it was necessary to add zero in the form of two contours $C_+$ and $C_-$ \eqref{addzero}.}
\label{fig:wplane}
\end{figure}
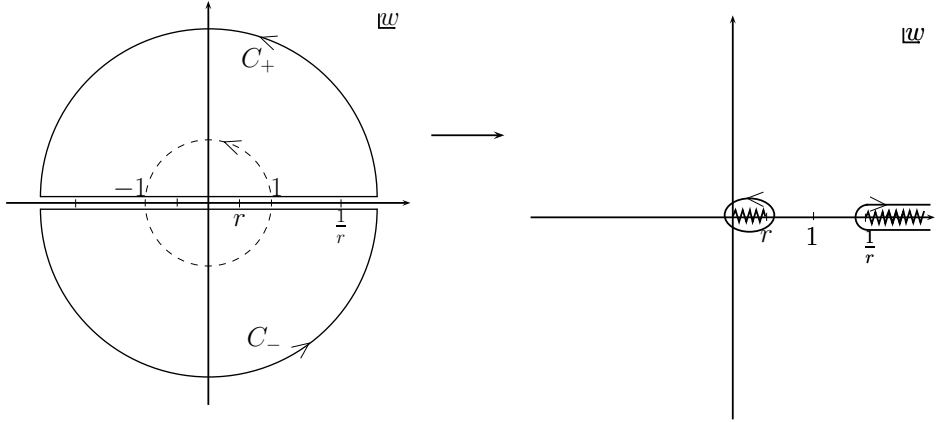

In this case,
\beq
\begin{split}
\wh{h}_1(s,w)=(-1)^{\frac{q}{2}-1} \frac{\Gamma(s+1) \Gamma\left(s+q-2\right)}{\Gamma\left(s+\frac{q}{2}-1\right)\Gamma\left(s+\frac{q}{2}\right)} \left(\wh{h}_2(s,w)+\wh{h}_3(s,w)\right)\,,\\
s=0,1,2\ldots\,, \quad q=2,4,6,\ldots \,.
\end{split}
\label{h123evenq}
\eeq
After plugging eq.~\eqref{h123evenq} in the Euclidean inversion formula \eqref{euclideaninv}, we can deform the contour towards the interior on $\wh{h}_2$ and towards the exterior on $\wh{h}_3$ (see fig.~\ref{fig:wplane}).
When deforming the contour towards the interior, divergences may arise when shrinking the circles around $w=0$ and $w=r$. The former is not an OPE limit. It lies at the boundary of the region of convergence of the bulk OPE, which is not positive. In the case of the Caron-Huot formula \cite{Caron-Huot:2017vep}, positivity was used to place a bound on the growth of the correlator in the Regge limit, which has the same role there as the small $w$ limit here. Deprived of this tool, we currently have no way to constrain the growth of the correlator in general. 
If for small $w$  the two-point function is bounded by a power, then for $s$ large enough the circle around $w=0$ can be shrunk. Concretely,
\beq
\mathrm{if} \;\; g(r,w) \lesssim w^{-s_{\star}}\,, \quad \text{as } w \to 0 \qquad \text{then the formula is valid for} \; s >s_{\star} \,,
\label{sstar}
\eeq
since the integrand in \eqref{euclideaninv} is then bounded by $w^{s-1-s_{\star}}$ for $w \to 0$. In all the perturbative examples of section \ref{sec:examples}, $s_\star=0$. We should also keep in mind that the inversion formula only converges after the light defect primaries have been subtracted  -- see the discussion below \eqref{Euclidstrip}. A defect block of transverse spin $s$ has $s_\star=s$: hence, unless special cancellations happen, the presence of primaries with $\Dhat<p/2$ and large $s$ worsens the convergence of the Lorentzian formula.

The point $w=r$ is a bulk OPE singularity, that includes power laws of the kind $(r-w)^{-\Delta_\phi+\tau/2}$. The integral converges for negative enough $\Delta_\phi$, and can then be analytically continued. The procedure is allowed because the angular integral in the original Euclidean formula is convergent for all values of $\Delta_\phi$.  Then we find that the OPE coefficient can be expressed in terms of the discontinuity of the two-point function across the branch cut running from $w=0$ to $w=r$: 
\beq
 \Disc \,g(r,w)= g(r,w+\ii 0)-g(r,w-\ii 0)\,.
\eeq
Note that for even $q$ the measure does not have any branch cuts.
Proceeding in the same way for $\wh{h}_3$, now deforming the contour towards the exterior, we obtain the following Lorentzian formula:
\begin{multline}
b(\Dhat,s)=\frac{1}{2} \frac{K_\Dhat}{K_{p-\Dhat}} (b_0(\Dhat,s)+b_\infty(\Dhat,s))\,, \\
\begin{aligned}
b_0(\Dhat,s)&=-\int\limits_0^1dr \int\limits_0^r \frac{dw}{\ii\pi w}
w^{2-q} (1-w^2)^{q-2} (1-r^2)^p r^{-p-1}\wh{h}_2(s,w)\Psi_\Dhat(r)\Disc\, g(r,w)\,, \\
b_\infty(\Dhat,s)&=\int\limits_0^1dr \int\limits_{1/r}^\infty\frac{dw}{\ii\pi w}
w^{2-q} (w^2-1)^{q-2} (1-r^2)^p r^{-p-1}\wh{h}_3(s,w)\Psi_\Dhat(r)\Disc\, g(r,w)\,.
\end{aligned}
\label{lorinveven}
\end{multline}
Note that in any theory with $q>2$ the two-point function obeys $g(r,w)=g(r,\tfrac{1}{w})$, and thus it follows that $\Disc\, g(r,w)=-\Disc\, g(r,1/w)$. In the case $q=2$ the symmetry of the two-point function is only present in a parity invariant theory, which is assumed here since we have taken the blocks in eq.~\eqref{eq:geta} to be symmetric under  $w \to \tfrac{1}{w}$. Therefore,
\beq
b_0(\Dhat,s)=b_\infty(\Dhat,s)~.
\eeq

\subsubsection*{\texorpdfstring{Odd $q$}{Odd q}}

When the codimension is odd, the relation \eqref{h123evenq} cannot be globally valid, because of the cuts in $\wh{h}_2$ and $\wh{h}_3$. However, we only need a relation which is valid upon integration. In other words, we can integrate zero in the form of a combination of $\wh{h}_2$ and $\wh{h}_3$ along the contours $C_+$ and $C_-$ in fig.~\ref{fig:wplane}.
In fact, we only integrate $\wh{h}_3$ which does not grow at infinity
\beq
b(\Dhat,s)=b(\Dhat,s)+c_+\oint_{C_+}dw\,[\dots] \wh{h}_3(s,w)+
c_-\oint_{C_-}dw\,[\dots] \wh{h}_3(s,w)\,.
\label{addzero}
\eeq
where the dots stand for everything in eq.~\eqref{euclideaninv} except $\wh{g}_s(\eta)$. We need linear combinations of the conformal block and the $\wh{h}$'s which does not blow up in $w=0$. This fixes
\beq
c_\pm=-e^{\pm\ii \frac{\pi}{2}q}\frac{\Gamma(s+1) \Gamma\left(s+q-2\right)}{\Gamma\left(s+\frac{q}{2}-1\right)\Gamma\left(s+\frac{q}{2}\right)}\,.
\eeq
Notice that the combination is different in the upper and lower plane, due to the cut in $\wh{h}_3$, which extends from $w=-1$ to $w=1$. The structure of cuts in the integrand is complicated by the contribution of $\mu(r,w)$ -- see eq.~\eqref{euclideaninv}. As an analytic function of $w$, $\mu(r,w)$ for odd $q$ has a cut running over the whole real axis. Indeed, let us start from $w$ purely imaginary. In this case
\begin{align}
\left|\frac{w}{2 \ii}-\frac{1}{2 \ii w}\right|^q &=\frac{e^{-\ii \frac{\pi}{2}q}}{2^q}\left(\frac{(w+1)(w-1)}{w}\right)^q, \qquad -\ii w>0\,, \\
\left|\frac{w}{2 \ii}-\frac{1}{2 \ii w}\right|^q &=(-1)^q\frac{e^{-\ii \frac{\pi}{2}q}}{2^q}\left(\frac{(w+1)(w-1)}{w}\right)^q, \qquad -\ii w<0\,.
\end{align}
So the discontinuity of the measure on the real axis is
\beq
\Disc\, \mu(u,w)=\mu(u,w+\ii 0)-\mu(u,w-\ii 0)=\left(1-(-1)^q\right)\mu(u,w+\ii 0),\qquad w \in  \mathbb{R} \,.
\eeq
We can now deform the contour of integration in $b(\Dhat,s)$ to the real axis, and drop the arcs at infinity in $C_\pm$. We first notice that the discontinuity in $\mu(r,w)$ offsets the difference between $c_+$ and $c_-$. Let us consider first the region $\abs{w}>1$. There the only further discontinuity comes from the correlator $g(r,w)$. In the complementary region $\abs{w}<1$, the integral in $b(\Dhat,s)$ on the r.h.s. of eq.~\eqref{addzero} must be taken into account. This combines with the contributions from $C_+$ and $C_-$ by virtue of the following relations:
\begin{align}
\wh{h}_1(s,w)-c_+ \wh{h}_3(s,w)&=-e^{-\ii\frac{\pi}{2}q}\frac{\Gamma(s+1) \Gamma\left(s+q-2\right)}{\Gamma\left(s+\frac{q}{2}-1\right)\Gamma\left(s+\frac{q}{2}\right)} \wh{h}_2(s,w)  \,, \qquad \Im w>0\,,\ \abs{\Re w}<1\,, \\
\wh{h}_1(s,w)-c_- \wh{h}_3(s,w)&=-e^{\ii\frac{\pi}{2}q}\frac{\Gamma(s+1) \Gamma\left(s+q-2\right)}{\Gamma\left(s+\frac{q}{2}-1\right)\Gamma\left(s+\frac{q}{2}\right)} \wh{h}_2(s,w) \,, \qquad \Im w<0\,,\ \abs{\Re w}<1\,.
\end{align}
Putting all together, we recover a result identical to the case of even codimension, namely eq.~\eqref{lorinveven}.

\subsubsection*{\texorpdfstring{Lorentzian inversion formula for codimension $q$}{Lorentzian inversion formula for codimension q}}

All in all, we obtain the following Lorentzian formula, valid for both even and odd $q$
\beq
b(\Dhat,s)=-\frac{K_\Dhat}{K_{p-\Dhat}}  \int\limits_0^1dr \int\limits_0^r\frac{dw}{\ii\pi w}
w^{2-q} (1-w^2)^{q-2} (1-r^2)^p r^{-p-1}\wh{h}_2(s,w)\Psi_\Dhat(r)\Disc g(r,w)\,.
\label{eq:inversionformula}
\eeq 
Finally, we can change coordinates to $z$ and $\bz$. If we are only interested in the poles of $b(\Dhat,s)$ in $\Dhat$ corresponding to the exchanged operators, and not their shadows, we need only keep the second addend in $\Psi_\Dhat(r)$ -- see eq.~\eqref{bdelta}. We find
\beq
\begin{split}
\left. b(\Dhat,s)\right|_{\textup{poles}}= \int\limits_0^1 \frac{dz}{2z} z^{-\frac{\ttr}{2}} 
\int\limits_1^{\frac{1}{z}}\! \frac{d\bz}{2 \pi \ii} \, &(1-z\bz)(\bz-z)\bz^{-\frac{\Dhat+s}{2}-2}\;
{}_2F_1\left(s+1,2-\frac{q}{2},\frac{q}{2}+s,\frac{z}{\bz}\right) \\
&\times{}_2F_1\left(1-\Dhat,1-\frac{p}{2},1+\frac{p}{2}-\Dhat, z\bz \right) \Disc\, g(z,\bz)\,,
\label{generating}
\end{split}
\eeq
where we have used hypergeometric identities to simplify the equation.
The cut between $w=0$ and $w=r$ has been mapped to the line $\bz\in [1,1/z]$, and can be computed by going around the branch point at $\bz=1$. Notice that, due to the inverse proportionality relation between $w$ and $\bz$ in eq.~\eqref{utoradial}, $\Disc\, g(r,w)=-\Disc\, g(z,\bz)$.
Eqs. \eqref{eq:inversionformula} and \eqref{generating} are analytic in $s$. However, we stress again that their validity cannot be established without knowledge of the behavior of $g(r,w)$ for $w \to 0$, or equivalently $w \to \infty$. If $s_\star<\infty$, $s_\star$ being defined in eq. \eqref{sstar}, 
the function $b(\Dhat,s)$ defined by eq. \eqref{eq:inversionformula} is identical to the function obtained via the Euclidean inversion formula eq. \eqref{euclideaninv} for all integer values of $s>s_\star$.
But now, analyticity in $s$ implies that the defect operators organize in analytic trajectories for $s> s_\star$.

Let us also note that, similarly to the formula obtained in \cite{Caron-Huot:2017vep}, the discontinuity in eq.~\eqref{generating} vanishes for a single defect block, and thus its validity cannot be verified term by term in a defect block decomposition. This is to be contrasted with the Euclidean formula \eqref{euclideaninv}, where the poles precisely arise order by order in the defect OPE expansion of the correlator, as we discussed around eq. \eqref{EucliDeltapoles}.

\medskip

Poles of $b(s,\Dhat)$ in $\ttr$ arise from the lower bound of integration in $z$, and we can study  eq.~\eqref{generating} in an expansion for small $z$,
\begin{align}
b(\Dhat,s)\big\vert_{\mathrm{poles}} &= \int\limits_0^1\frac{dz}{2z} z^{-\frac{\ttr}{2}}\sum\limits_{m=0} z^m \sum\limits_{k=-m}^{m}c_{m,k}(\Dhat,s) B(z,\beta+2k)\,,\quad
B(z,\beta) &\colonequals \int\limits_1^\infty \frac{d\bz}{2 \pi \ii } \bz^{-\frac{\beta }{2}-1} \Disc\, g(z,\bz) \,,
\label{eq:smallzgen}
\end{align}
where  $\beta=\Dhat+ s$, and where $c_{m,k}(\Dhat,s)$ are trivially obtained from the $z$ expansion of the integrand in \eqref{generating}, with $c_{0,0}(\Dhat,s)=1$.
Note that in eq.~\eqref{eq:smallzgen} we pushed the upper bound of the $\bz$ integration to infinity, which will not modify the poles of $b(s,\Dhat)$ in $\Dhat$, \emph{provided} $g(z,\bz)$ behaves as \eqref{sstar}.
This follows from the behavior of \eqref{eq:smallzgen} for small $z$ and with $\bz \sim \tfrac{1}{z}$.\footnote{The upper bound of the $\bz$ integration can only produce poles in $s$, and provided $g(z,\bz)$ grows as given in eq.~\eqref{sstar} for $w \to 0$, then these poles will appear only for $s \leqslant s_\star$, that is for $s$ outside the range of applicability of the formula.}

In a series expansion for small $z$, the functions $B(\beta,z)$ will give the following contributions to \eqref{eq:smallzgen}:
\beq
\sum\limits_{m=0} z^m \sum\limits_{k=-m}^{m}c_{m,k}(\Dhat,s) B(z,\beta+2k)=\sum\limits_m b_m(\Dhat,s) \; z^{\tfrac{\ttr_m(\beta,s)}{2}}\,,
\label{eq:Bstruct}
\eeq
with each term producing a pole for $\ttr = \ttr_m(\beta,s)$ in $b(\Dhat,s)$, signaling a defect operator with that transverse twist.
The OPE coefficients are obtained from the $\Dhat$-residue of $b(s,\Dhat)$, at fixed $s$, according to \eqref{eq:resb}, and so they are obtained from the coefficients in \eqref{eq:Bstruct} after correcting by a Jacobian factor as
\beq
\label{eq:Jac}
b^2_{s,\Dhat}= \left(1- \frac{d\ttr_m(\beta,s)}{d\beta}\right)^{-1} b_m(\Dhat,s) \Big\vert_{\beta=\ttr_m(\beta,s)+2s}\,.
\eeq


\subsection{Contributions from a single bulk block}
\label{eq:CH_singlebulk}

Of course in general one does not have access to the full two-point function. Representing the two-point function by its bulk OPE, we now discuss what can be inferred from knowledge of individual exchanged bulk blocks, making contact with the considerations of section \ref{sec:lightdefect}.

As discussed in section \ref{sec:lightdefect}, knowledge of the low twist operators appearing in the bulk OPE translates into statements about the large transverse spin defect spectrum. The analysis of section \ref{sec:lightdefect} is not free from assumptions, similarly to the usual lightcone story applied to the four-point function of local operators. In the latter case, only recently have some of the assumptions started to be put on a firmer footing \cite{Qiao:2017xif}. 
We can now recover the results of section \ref{sec:lightdefect} making use of the inversion formula. \emph{Assuming} that the correlation function behaves as in eq.~\eqref{sstar}, the inversion formula shows that operators organize in analytic families for $s> s_\star$.
Now we can prove that these trajectories have accumulation points for $s\to \infty$ at $\ttr \to \Delta_{\phi} + 2m$.
Furthermore, unlike in section \ref{sec:lightdefect}, where we obtained the contribution of an exchanged bulk block to the defect spectrum in a $1/s$ expansion, the results obtained through \eqref{eq:inversionformula} amount to the full contribution of the block at finite $s > s_\star$. These results therefore resum the $1/s$ expansion that could be obtained from carrying out the procedure of section \ref{sec:lightdefect} to all orders in $1/s$, as done in \cite{Alday:2015ewa,Alday:2015ota,Alday:2016jfr,Simmons-Duffin:2016wlq} for the case without defects.

Let us see how the transverse derivatives come about in this context. 
For large transverse spin $s$, we see from \eqref{eq:smallzgen} that the integral is dominated by $\bz \to 1$. 
From the behavior of the bulk blocks in this limit \eqref{singlecoll}, we find that the leading contributions come from operators with lowest twist, $\tau=\Delta-J$, which contribute as
\beq 
g(z,\bz) \sim (1-\bz)^{-\Delta_\phi + \frac{\tau}{2}} \left(\textup{a function of }z\right)+ \ldots \,, \qquad \text{for } \bz \to 1\,.
\label{eq:zbto1ofg}
\eeq
The leading contribution is always the identity. As it is expected and we confirm below, the inversion of the identity yields the spectrum of the trivial defect, \ie, the transverse derivatives and their OPE coefficients.

Note that we can perform the $\bz$ integral that defines $B(z,\beta)$ in eq.~\eqref{eq:smallzgen}, \emph{at fixed $z$}, block by block. Indeed, the bulk channel OPE still converges in the whole region $0<z<1/\bz<1$. This is most easily seen in the coordinates defined in appendix \ref{sec:rhobulk}, to which we refer.
However, it will be convenient to work in a small $z$ expansion, which is where the poles of $b(\Dhat,s)$ arise from, and so we now discuss the circumstances under which this is allowed.

\subsubsection*{Small $z$ expansion}

The small $z$ expansion does not commute with the infinite sum over bulk blocks, as it is clear from the fact that while \eqref{eq:smallzgen} should behave like \eqref{eq:Bstruct}, all the blocks except the identity contain a single logarithm of $z$ as $z\to0$, see eq. \eqref{doublecollall}.\footnote{One might complain that using eq. \eqref{doublecollall} of the block is not allowed here because the lightcone expansion of the block does not converge in the whole range $1<\bz<\infty$. One can instead use the expansion presented in section 4.2.1 of \cite{Billo:2016cpy}, where the expansion parameter is $\frac{1-\bz}{\bz}$, which can be integrated in the desired range. The result still contains a single $\log z$.}
This is exactly the same problem discussed in \cite{Caron-Huot:2017vep} for the four-point function without defects. There, in section 4.3.2, a way out was found -- see also \cite{Simmons-Duffin:2016wlq}: after subtracting a known sum from the inversion formula, one can commute the small $z$ limit with the block expansion. In this section, we will content ourselves of computing the contributions of individual bulk blocks after taking their small $z$ limit, without any subtraction. We expect that the error we make becomes small when the defect spectrum differs from the one of the trivial defect by small anomalous dimensions: indeed, in this case also the r.h.s. of eq. \eqref{eq:Bstruct} is well approximated by an expansion up to a single $\log z$. This happens for instance at large spin, where the analytic functions in $s$ that we find below resum part of the lightcone expansion. In some specific situations, the result is actually exact down to $s=s_\star$. These are the cases considered in section \ref{sec:examples}: defects whose deviation from the trivial one is controlled by a perturbative parameter. At leading order in this coupling, a single logarithm of $z$ is all what there is on the r.h.s. of eq. \eqref{eq:Bstruct}. Furthermore, in these examples a finite number of bulk channel blocks have a non-zero discontinuity, therefore we are free to take small $z$ block by block.\footnote{A similar situation was also very recently discussed for the four-point function case in \cite{Alday:2017zzv}.} 

It should be borne in mind that, to go beyond these results, a procedure similar to \cite{Caron-Huot:2017vep} is needed.

\subsubsection*{Identity exchange and the transverse derivative operators}

As discussed above, the leading contributions to the large transverse spin spectrum come from leading twist bulk operators, and thus the identity operator, which has twist zero, dominates.\footnote{All remaining operators are constrained by unitarity bounds to have $\tau=\Delta- J>0$, provided $d>2$ which is assumed throughout this work since we consider $q>1$.}
If this is the only exchanged bulk operator then from \eqref{eq:crossingsy} we find $s_{\star}=-\Delta_\phi$, and the inversion formula \eqref{eq:inversionformula} is valid for all spins starting at $s=0$. This happens for the trivial defect \eqref{eq:trivialdef}, and thus we recover the full spectrum. However, if the identity is just part of a more complicated two-point function the $w \to 0$ behavior, and thus $s_\star$ can be modified (see, \eg, the simple example in subsection \ref{subsubsec:free}).

Taking the leading small $z$ term in the identity contribution to $g(z,\bz)$ we find
\beq
B(z,\beta) = \int\limits_{1}^{\infty} d \bz \bz^{-\frac{\beta}{2}-1} \frac{1}{2 \pi \ii }\Disc\left(\frac{(1-\bz)}{\sqrt{z \bz}}\right)^{-\Delta_\phi} = \frac{z^{\frac{\Delta \phi}{2}}\Gamma\left(\frac{\Dhat+s+\Delta_{\phi}}{2}\right)}{\Gamma \left(\frac{\Dhat+s-\Delta_{\phi}}{2}+1\right) \Gamma \left(\Delta_{\phi}\right)}\,,
\label{eq:idinv}
\eeq
where we see that in \eqref{eq:smallzgen} this produces a pole in $\ttr= \Delta_\phi$, corresponding to the leading twist defect primary operator. The residue of $B(z,\beta)$ matches precisely with the OPE coefficient of the trivial defect \eqref{btrivial}, for $m=0$.
Subleading powers in the small $z$ expansion of the identity block and of \eqref{generating} produce poles corresponding to the rest of the trivial spectrum, \eqref{trivialspectrum} with $m>0$.

Since the bulk identity exchange corresponds to the leading contribution to the spectrum at large $s$, we thus recover the existence of transverse derivative operators with $\ttr\to\Delta_\phi + 2m$ as $s \to \infty$. A main difference with respect to section \ref{sec:lightdefect} is that now we obtain the full OPE coefficient \eqref{btrivial}, instead of an asymptotic series in $1/s$.

Note that the integral in \eqref{eq:idinv} naively diverges for large $\Delta_\phi$, but the result can be defined by analytic continuation and is finite, similarly to  what was observed in \cite{Caron-Huot:2017vep}.\footnote{The result of the Euclidean inversion formula gives a finite answer that is analytic in $\Delta_\phi$. For $\Delta_\phi < 1$ eq.~\eqref{eq:idinv} converges and thus it matches the result of the Euclidean inversion. The integral in \eqref{eq:idinv} can then be analytically continued from there to $\Delta_\phi \geqslant 1$. This will also happen for the exchange of low dimensional bulk primaries, as the behavior for $\bz \to 1$ of $g(z,\bz)$ is controlled by the bulk channel OPE -- see \eqref{eq:zbto1ofg}. For low dimensional bulk blocks then the result should also be obtained by analytic continuation.} 

Finally we note that $\ii \, \Disc g(z,\bz)$ in \eqref{generating} does not have a definite sign, in contrast to the case of the double-discontinuity in \cite{Caron-Huot:2017vep}.
This is clear from the identity contribution in \eqref{eq:idinv} where
\beq
\left(2 \ii\right)^{-1} \Disc\left((1-\bz)^{-\Delta_\phi}\right) = (\bz-1)^{-\Delta_\phi } \sin(\pi  \Delta_\phi )\,.
\eeq 
Even though positivity of the defect OPE coefficients requires the residues of $b(\Dhat,s)$ in \eqref{generating} to have definite sign, as is the case for \eqref{eq:idinv} above, this does not follow form the sign of the discontinuity.

\subsubsection*{Leading bulk twist contribution}

The defect operator dimensions and OPE coefficients obtained from the inversion of the identity block will then be corrected for finite spin by the presence of all the remaining bulk blocks.
We define the anomalous dimension of the transverse derivative operators whose dimensions approach $\Delta_\phi + 2m$ as
\beq
\gamma_{s,m} \colonequals  \ttr_{m}-(\Delta_\phi+2m)\,.
\label{eq:gammasdef}
\eeq
As discussed above, if the $\gamma_{s,m}$ are small then we can consider the the small $z$ limit of the bulk block decomposition.

We can draw from \eqref{eq:zbto1ofg} a first general observation: if the exchanged operator has twist $\tau= 2 \Delta_\phi+2n$, with $n \geqslant 0$ an integer, the contribution of the relative block to the discontinuity of $g(z,\bz)$ vanishes. In other words, exact double twists of the external operator have zero discontinuity and do not contribute under the inversion formula. Note that while the discontinuity naively vanishes also for negative integer $n$, the integral is divergent for $\bz \to 1$ in this case. One must then first compute the discontinuity for arbitrary $n$ and perform the integration. In the end, when $n$ is taken to be a negative integer, the zero of the discontinuity cancels the divergence in the integral, and the final result of the inversion formula is finite. This is in precise agreement with the results of section \ref{subsec:transder}: the bulk blocks with non vanishing discontinuity either give singular contributions to $g(z,\bz)$ as $\bz \to 1$, or contributions that can be made singular by acting with the Casimir.\footnote{The same behavior is observed for the inversion formula of the four-point function with no defects as pointed out in \cite{Caron-Huot:2017vep} -- see also \cite{Alday:2017vkk}.}

Then, the contribution of the identity and a bulk primary $O$ of twist $\tau$ and spin $J$ to \eqref{eq:Bstruct} has the following form in a small $z$ expansion:
\beq
\label{applGen}
\sum\limits_{m=0} I_{m}(\Dhat,s)z^{\frac{\Delta_{\phi}}{2}+m}+
c_{\phi\phi O}a_{O}  \sum\limits_{m=0}\left(C_{1}^{m}(\Dhat,s)+C_{2}^{m}(\Dhat,s)\log{z}\right)z^{\frac{\Delta_{\phi}}{2}+m}\,.
\eeq
The leading contributions to the anomalous dimension, and the correction to the OPE coefficients \eqref{btrivial} are obtained from the above as\footnote{In case $I_m$ is zero the denominator should be the first non-zero order, this happens for instance if the external operator $\phi$ is perturbatively close to the unitarity bound, since in this case $b_{s,m}^2=0$ for $m\neq0$ and $\Delta_\phi=\tfrac{d}{2}-1$.}
\beq
\gamma_{s,m}=\frac{2c_{\phi\phi O}a_{O}C_{2}^{m}}{I_{m}}\Bigg\vert_{\Dhat=\Delta_{\phi}+s+2m}\,,
\quad
\delta b^2_{s,m}=\frac{d\gamma_{s,m}}{d\Dhat} I_{m}+\gamma_{s,m}\frac{dI_{m}}{d\Dhat}+c_{\phi\phi O}a_{O}C_{1}^{m}\Big\vert_{\Dhat=\Delta_{\phi}+s+2m} \,. 
\label{lianom}
\eeq 
Applying these results to the bulk collinear block given in eq.~\eqref{singlecoll} and expanding the answer for large $s$, we have recovered the results obtained with the lightcone approach of section \ref{sec:lightdefect}, for different values of $m$ and to second order in $\tfrac{1}{s}$. To this order, the Jacobian factor contribution in \eqref{lianom} is crucial.
 
This proves the existence of the individual transverse derivative operators, instead of the averaged statement obtained with the lightcone analysis.\footnote{To be precise, for a given finite spin, it may happen that the contributions from the various bulk primaries to the residue of a certain pole sum up to zero.  However, this cannot happen for sufficiently large spin, where the corrections from different exchanged operators are of different size. In this sense our results, similarly to those of \cite{Caron-Huot:2017vep}, establish the existence of each individual transverse derivative operator for sufficient large $s$.}

While the methods of section~\ref{sec:lightdefect} only provide an asymptotic series in $\tfrac{1}{s}$, the inversion formula yields the contribution of a given bulk block to the anomalous dimension and OPE coefficient of a defect operator of any transverse spin $s> s_\star$. As an example, we compute the full correction arising from the exchange of a bulk scalar, and from the bulk stress tensor, to the leading twist defect operator $\gamma_{s,0}$.

\subsubsection*{Scalar operator exchange}

Let us first obtain how a scalar operator $O$ of dimension $\Delta$ contributes to $\gamma_{s,0}$.
Apart from the scalar operator $O$, here we only take into account the contribution of the identity.
While the bulk blocks are not known in closed form, we can make use of the representation of the scalar block as an infinite sum of hypergeometric functions as given in appendix~B of \cite{Billo:2016cpy}.\footnote{Alternatively we could have used the recursion relation for the bulk blocks obtained in \cite{Billo:2016cpy}.} Taking the leading $z \to 0$ term of the block we apply \eqref{eq:smallzgen} term by term in the block representation as an infinite sum.
This amounts to a representation of the block as an infinite sum in powers of $\tfrac{\bz-1}{\bz}$ that converges for all of $\bz$ in the integration region of \eqref{eq:smallzgen}. We then commute the integral over $\bz$ with the infinite sum, and are able to resum the result to find
\beq
\gamma_{s,0}\big\vert_{\Delta, J=0}= -c_{\phi \phi O }a_{O}\frac{2^{\Delta } \Gamma \left(\frac{\Delta +1}{2}\right) \Gamma (\Delta_\phi ) \Gamma (s+1) \, _3F_2\left(\frac{\Delta-q+2 }{2},\frac{\Delta }{2},\frac{\Delta -2\Delta_\phi +2 }{2};\frac{\Delta }{2}+s+1,\Delta -\frac{d-2}{2};1\right)}{\sqrt{\pi } \Gamma \left(\frac{\Delta }{2}\right) \Gamma \left(\Delta_\phi -\frac{\Delta }{2}\right) \Gamma \left(\frac{\Delta }{2}+s+1\right)}\,.
\label{gammaScalar}
\eeq
We can proceed similarly for subleading transverse twists by keeping more terms in the small $z$ expansion, but since the resulting anomalous dimension have longer expressions we do not display them here.
Note that by taking the $z\to0$ limit of the scalar block we are assuming small anomalous dimensions, and the result we present here should be seen as the leading contribution in the small parameter that controls the anomalous dimension.

\subsubsection*{Stress tensor exchange}

We now turn to the stress tensor exchange, whose blocks are easily computed by solving the recursion relation obtained in \cite{Billo:2016cpy} for an operator with $J=2$ and $\Delta=d$.
In this case it is easy to obtain a closed form answer for the $z \to 0$ limit of the block, and again we compute its contribution to $b(s,\Delta)$ via \eqref{eq:smallzgen}.
Again considering only the stress tensor and the identity, we obtain the following anomalous dimension:
\beq
\gamma_{s,0}\big\vert_{\Delta=d, J=2}=
-c_{\phi \phi T}a_{T} \frac{2^{d} \Gamma \left(\frac{d +3}{2}\right) \Gamma (\Delta_\phi ) \Gamma (s+1) \Gamma \left(\Delta_\phi +s -\frac{p}{2}\right)}{\sqrt{\pi } \Gamma \left(1+\frac{d }{2}\right) \Gamma \left(\Delta_\phi -\frac{d-2}{2}\right) \Gamma \left(\frac{q}{2}+s\right) \Gamma (\Delta_\phi + s )}\,.
\label{gammaStress}
\eeq

\bigskip

In section \ref{sec:examples} we will only need the contributions \eqref{gammaScalar} and \eqref{gammaStress}, but of course one can repeat this procedure for subleading twists.
Similarly the corrections to the OPE coefficients of the trivial defect \eqref{btrivial} arising from either of these two exchanges can be computed. Since their expressions are not particularly illuminating, and in the case of the scalar exchange the infinite sums were not performed to get a closed form result, we do not present them here. Note that in computing the OPE coefficient one must include the Jacobian factor of eq.~\eqref{eq:Jac}.

\subsubsection*{Behavior of a single bulk block  as $w \to 0$}

While we cannot bound the growth of $g(z,\bz)$ as $w \to 0$, we can check the behavior of a single bulk block. This is trivial for the cases in which the blocks are known in closed form, and one finds that for codimension two and $d=4,6$, the behavior is power-law. \emph{Assuming} that the behavior of the bulk block is power-law for all values of $p$ and $q$ we can solve both the quadratic and quartic Casimir equations in the small $w$ limit to find\footnote{There is another nontrivial solution allowed by the Casimir that does not match the behaviors obtained in codimension two.}
\beq 
f_{\Delta,J} \sim w^{-\frac{p}{2}} f(r)\,, \qquad \text{as} \qquad w \to 0\,,
\label{smallwblock}
\eeq
where $f(r)$ is a function of $r$ that is fixed up to two constants by the Casimir equations. Putting in the behavior of the prefactor in \eqref{eq:crossingsy} we find that for a single bulk block $s_\star=\tfrac{p}{2}-\Delta_\phi$. Indeed, the anomalous dimensions in eqs. \eqref{gammaScalar} and \eqref{gammaStress} become singular precisely at $s=s_\star$. Of course, the small $w$ behavior can be different from eq. \eqref{smallwblock}, for a theory where an infinite number of bulk blocks is exchanged.

\section{Defect operators and motions in AdS}
\label{sec:ads}

\begin{figure}
\begin{center}
\scalebox{0.65} 
{
\begin{pspicture}(0,-3.405169)(20.089062,4.3872647)
\definecolor{color818}{rgb}{0.6,0.6,0.6}
\definecolor{color824}{rgb}{1.0,0.2,0.2}
\definecolor{color830b}{rgb}{0.6078431372549019,0.6078431372549019,0.6078431372549019}
\rput{-90.85189}(10.454097,11.828885){\psellipse[linewidth=0.04,linecolor=color818,linestyle=dotted,dotsep=0.16cm,dimen=outer](11.054201,0.76453924)(0.623327,1.9772725)}
\psellipse[linewidth=0.04,dimen=outer](11.029267,0.78941053)(0.6233276,1.9772724)
\pscircle[linewidth=0.04,linecolor=color818,dimen=outer](11.036735,0.7894106){2.0145795}
\psdots[dotsize=0.13878673,linecolor=color824](13.049987,0.7548311)
\psdots[dotsize=0.13878673](10.509987,1.7348443)
\psdots[dotsize=0.13878673,linecolor=color824](9.049987,0.7548311)
\psdots[dotsize=0.36002687,dotstyle=x](11.609986,1.314831)
\psdots[dotsize=0.36002687,dotstyle=x](10.469987,0.19483104)
\rput{-186.66972}(20.646276,1.9103904){\pstriangle[linewidth=0.04,linecolor=color830b,dimen=outer,fillstyle=solid,fillcolor=color830b](10.267479,1.4567261)(0.22,0.2)}
\psline[linewidth=0.04cm,linecolor=color830b](10.354203,2.2334206)(10.282192,1.6176168)
\psline[linewidth=0.04,fillstyle=solid,fillcolor=black](15.55,3.9148312)(15.55,-2.485169)
\psline[linewidth=0.04,fillstyle=solid,fillcolor=black](19.55,3.9148312)(19.55,-2.485169)
\psellipse[linewidth=0.04,dimen=outer](17.55,3.9148312)(2.0,0.4)
\psellipse[linewidth=0.04,dimen=outer](17.55,-2.485169)(2.0,0.4)
\rput{192.96402}(36.449535,2.8947566){\psarc[linewidth=0.04](18.389214,-0.6232763){1.1564586}{126.815796}{168.3994}}
\rput{172.64798}(33.21903,-0.64210105){\psarc[linewidth=0.04](16.63014,0.7460547){3.3918831}{79.66636}{149.48438}}
\rput{344.29837}(1.4150791,4.4529448){\psarc[linewidth=0.04,linestyle=dashed,dash=0.16cm 0.16cm](16.854656,-2.9048398){3.9400046}{73.564606}{124.446365}}
\rput{192.96402}(36.449535,2.8947568){\psarc[linewidth=0.04,linestyle=dashed,dash=0.16cm 0.16cm](18.389215,-0.62327623){1.1564587}{177.44394}{221.3533}}
\rput{192.96402}(35.72716,9.252683){\psarc[linewidth=0.04](18.389214,2.5967238){1.1564586}{126.815796}{168.3994}}
\rput{192.96402}(35.731647,9.213192){\psarc[linewidth=0.04,linestyle=dashed,dash=0.16cm 0.16cm](18.389215,2.5767238){1.1564587}{177.44394}{221.3533}}
\rput{344.29837}(0.56138736,4.5816884){\psarc[linewidth=0.04,linestyle=dashed,dash=0.16cm 0.16cm](16.894657,0.25516015){3.9400046}{73.564606}{124.446365}}
\psline[linewidth=0.04,linecolor=color818,arrowsize=0.05291667cm 2.0,arrowlength=1.4,arrowinset=0.0,fillstyle=solid,fillcolor=black]{<-}(1.25,4.2948313)(1.25,-2.905169)
\psline[linewidth=0.04,linecolor=color818,fillstyle=solid,fillcolor=black](1.05,1.8948311)(1.45,1.8948311)
\psline[linewidth=0.04,linecolor=color818,fillstyle=solid,fillcolor=black](1.05,-0.505169)(1.45,-0.505169)
\psline[linewidth=0.04,linecolor=color824,fillstyle=solid,fillcolor=black](3.77,4.254831)(3.77,-2.145169)
\psline[linewidth=0.04,fillstyle=solid,fillcolor=black](2.57,3.9148312)(2.57,-2.485169)
\psline[linewidth=0.04,fillstyle=solid,fillcolor=black](6.57,3.9148312)(6.57,-2.485169)
\psellipse[linewidth=0.04,dimen=outer](4.57,3.9148312)(2.0,0.4)
\psellipse[linewidth=0.04,dimen=outer](4.57,-2.485169)(2.0,0.4)
\psellipse[linewidth=0.04,linestyle=dotted,dotsep=0.16cm,dimen=outer](4.57,1.914831)(2.0,0.4)
\psellipse[linewidth=0.04,linestyle=dotted,dotsep=0.16cm,dimen=outer](4.57,-0.48516896)(2.0,0.4)
\psline[linewidth=0.04,linecolor=color824,fillstyle=solid,fillcolor=black](5.77,3.614831)(5.77,-2.785169)
\psdots[dotsize=0.36002687,dotstyle=x](5.3699865,2.274831)
\psdots[dotsize=0.36002687,dotstyle=x](3.1699865,-0.76516896)
\rput{172.64798}(33.631073,5.771426){\psarc[linewidth=0.04](16.63014,3.9660547){3.3918831}{79.66636}{149.48438}}
\rput(0.57453126,-0.51516896){\Large $0$}
\rput(0.48453125,1.8648311){\Large $\pi$}
\rput(0.50453126,4.144831){\Large $\tau$}
\rput(13.114531,2.464831){\Large $\psi=\pi$}
\rput(8.9045315,2.414831){\Large $\psi=0$}
\rput(9.964531,2.004831){\Large $\varphi$}
\rput(19.20453,-3.0551689){\Large $\varphi$}
\psline[linewidth=0.04cm](4.55,-2.445169)(5.75,-2.765169)
\psline[linewidth=0.04cm](3.17,-2.765169)(4.57,-2.445169)
\rput(5.833906,-3.335169){\Large $\theta=\frac{\pi}{2}$}
\rput(3.2139063,-3.235169){\Large $\theta=0$}
\rput(13.7617185,0.72483104){\Large $\theta=\frac{\pi}{2}$}
\rput(8.35,0.72483104){\Large $\theta=\frac{\pi}{2}$}
\rput(11.087656,-1.695169){\Large $\theta=0$}
\end{pspicture} 
}
\end{center}
\caption{A three dimensional version of the configuration discussed in the text. In red, the line defect marks two lines along $\tau$ at $\theta=\frac{\pi}{2}$ and $\psi=0,\ \pi$. The particle spins fast along the $\varphi$ direction.}
\label{fig:aldaym}
\end{figure}
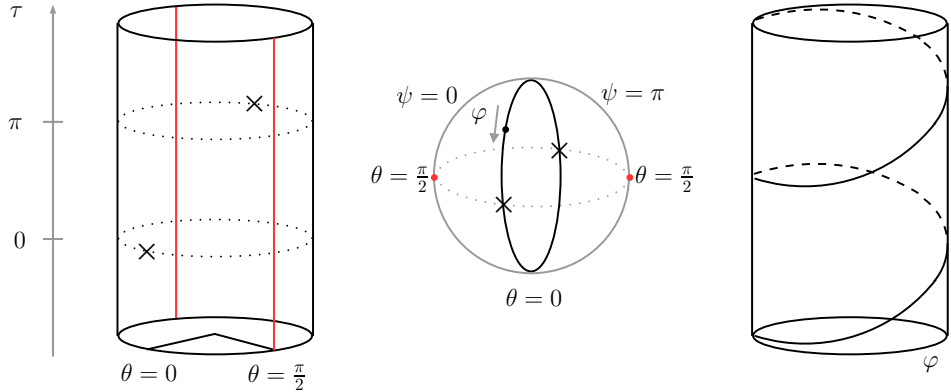

In sections \ref{sec:lightdefect} and \ref{sec:inversion} we discovered that transverse derivative operators are a necessary ingredient to ensure crossing symmetry of a correlator. In this section we give them a more physical characterization, which exploits the presence of a semi-classical limit of the states in radial quantization at large $s$ \cite{Alday:2007mf}. One of the main advantages of a picture directly based on the Hilbert space is that it provides intuition on the presence of other accumulation points in the spectrum. In particular, this analysis will allow us to identify another class of defect operators, which are present on the spectrum of Wilson lines in large $N$ gauge theories, similar in nature to the single-trace operators of the ambient CFT. 

Historically, the existence of double-twist primaries in the spectrum of an arbitrary CFT was first established precisely by looking at semi-classical states on a sphere. We give a brief review of the argument in \cite{Alday:2007mf}, before explaining its consequences for a defect CFT. Consider first the state created by a descendant of a low-lying scalar primary, schematically $\pa^\ell \phi$. At large spin, the state becomes classical, and can be approximated by a single particle which rotates close to the speed of light on the equator. Following \cite{Alday:2007mf}, we consider the four-dimensional case -- the generalization being straightforward. We parametrize the conformal time with $\tau$ and the 3-sphere with Hopf angles $(\theta,\varphi,\psi)$, with $\theta\in [0,\pi/2]$ and $\varphi$ and $\psi$ describing circles:
\beq
ds^2=-d\tau^2+d\theta^2+\cos^2\theta\, d\varphi^2+\sin^2\theta\, d\psi^2~.
\eeq 
The trajectory of the particle can be chosen to be $\tau=\varphi$, with $\theta=0$. For reasons to become clear shortly, it is convenient to apply a Weyl transformation to the cylinder, and turn it into the manifold $AdS_3\times S^1$. 
The authors of \cite{Alday:2007mf} chose the following parametrization in terms of coordinates $(u,\chi,\sigma,\psi)$:
\beq
ds^2=-du^2+d\chi^2-2 \sinh 2\sigma\,dud\chi+d\sigma^2+d\psi^2,
\label{metricadss1}
\eeq
where
\begin{subequations}
\begin{align}
\sinh\rho &= \frac{1}{\tan \th}~, \\
\sinh 2\si   &=-\sin(\tau-\varphi) \sinh 2\rho ~,   \\
e^{4\ii u}   &= e^{2\ii (\tau+\varphi)}
\frac{\cos(\tau-\varphi)+\ii \cosh 2\rho \sin (\tau-\varphi)}{\cos(\tau-\varphi)-\ii \cosh 2\rho \sin (\tau-\varphi)}~,    \\
\sinh 2\chi &= \frac{\cos(\tau-\varphi) \sinh 2 \rho}{\sqrt{1+\sin^2(\tau-\varphi) \sinh^2 2\rho}}\,.
\end{align}
\label{changeuschi}
\end{subequations}
This coordinate system has three key features. Firstly, the Hamiltonian equals the twist, \emph{i.e.} $\ii \pa_u=\ii \pa_\tau-(-\ii \pa_\varphi)$. Secondly, the fast particle on the sphere now sits at $\chi=+\infty$ and $\si=0$. Finally, the killing vector $S=-\ii \pa_\varphi \sim e^{2|\chi|}$ for large $\chi$ -- the explicit expression can be found in \cite{Alday:2007mf}. This construction does not tell us anything new about the single particle state: of course, the energy of the particle at rest in this two-dimensional quantum system is finite, which we knew from the start, since $\Delta-S=\Delta_\p$ for this state. However, consider now the state on the sphere in which two fast rotating particles are present, the second trajectory being $\tau=\varphi+\pi$. This state certainly exists, but now its scaling dimension is not obvious, since the theory is in general strongly coupled. However, in $AdS$ the particles sit infinitely far apart at $\chi=\pm \infty$, and the system is gapped due to the unitarity bound, so that no interaction is left in the infinite spin limit. The authors of \cite{Alday:2007mf} concluded that the the twist of such state is $\tau_\textup{DT}=\tau_{1}+\tau_{2}$, where $\tau_i$ is the individual twist of the operators responsible for the single particle states. This establishes the existence of double-twist operators.

Let us now consider a defect CFT. Since we aim at studying the scaling dimension of defect operators, we center the radial quantization in a point on the defect. On the cylinder, then, the defect extends along the time direction. We still consider a defect of codimension at least two, so we can localize it at $\th=\pi/2$. Depending on the dimension, it will cross the $\psi$ circle in two points, or it will fill it. In this setup, we reconsider the state of a particle rotating fast along the $\varphi$ circle. This is now a state with large transverse spin $s$. When we move on to $AdS_3\times S^1$, we find the defect at $\chi=\si=0$. The same argument as before allows us to conclude that, in the limit $s\to\infty$, the transverse twist of this state equals the twist of the bulk single particle state:
\beq
\ttr_\textup{TD}=\tau_\phi,\qquad s=\infty\,.
\eeq
We rediscover in this way the transverse derivative operators. This picture allows us to discuss the finite spin corrections as well, again following \cite{Alday:2007mf}. When the transverse spin is large but not infinite, we can use the mentioned behavior of the spin generator $S\sim e^{2|\chi|}$ to argue that the wave function of the particle is peaked at a distance from the defect of order $\frac{1}{2}\log s$. The particle is still localized also in $\sigma$, because the warp factor in the metric \eqref{metricadss1} favors the position $\sigma=0$. Since the system is gapped, the leading correction to the twist is a Yukawa potential, due to the exchange between the defect and the light particle of the leading twist state. We find therefore the correction
\beq
\ttr_\textup{TD}=\tau_\phi+\textup{const}\, e^{-\tau_\textup{min} \D\chi}=\tau_\phi +\textup{const}\, s^{-\tau_\textup{min}/2}\,.
\label{TDyukawa}
\eeq
Of course, we also know the precise coefficient from eq. \eqref{cmin}. Recall that, in  the case of the double-twist operators, the exponent of the correction is twice as large in absolute value: in AdS, this is simply a consequence of the different distance along $\chi$ of the interacting particles in the two cases.

In a gauge theory, one can also consider states in which the rotating particles are charged, and the authors of \cite{Alday:2007mf} consider this situation as well. In the case of a two-particle state, color flux extends between the two particles. The warp factor confines the flux close to $\sigma=0$, so that the twist of the state now equals the mass of a meson whose constituents are two light quarks. The flux-tube contributes to the mass with a constant energy per unit length. In the case of adjoint quarks, in $\mathcal{N}=4$ SYM the energy density is traditionally denoted $f(\lambda)$, and we keep this notation, even if the argument is valid in general. Since $\D\chi\simeq \log \ell$, the twist of the state is dominated by this contribution at large spin. Single trace operators precisely exhibit this behavior:
\beq
\tau_\textup{S.T.} \simeq f(\lambda)\log \ell\,.
\label{STtwist}
\eeq
In advocating this picture, we disregarded the possibility that the flux tube breaks: string-breaking effects are suppressed at large $N$, or in perturbation theory even at finite $N$, and we restrict our considerations to those cases. 

If we now consider the defect spectrum on a Wilson line, a very similar class of states emerges. At large $N$, these are created by the insertion of an adjoint operator $D^s\phi$ in the trace, and correspond to a color particle rotating around the defect. In AdS, we now see that this state is a meson whose constituents are a heavy quark sitting at the origin in the $(\chi,\sigma)$ plane, and a light quark placed far away along the $\chi$ direction, as before. Again, the distance between the two sources is half of what it was in the case of a single trace operator. We predict therefore the existence of defect operators of transverse twist
\beq
\ttr\simeq f(\lambda)\D\chi=\frac{f(\lambda)}{2} \log s\,,
\label{singletracedefect}
\eeq
where $f(\lambda)$ is the same function that appears in the anomalous dimension of single trace operators \eqref{STtwist}. It would be interesting to check this prediction, perhaps along the lines of \cite{Giombi:2017cqn}.

\section{Examples}
\label{sec:examples}

In this section we present a few illustrative examples. We start from a defect in free theory, which provides a simple instance where the inversion formula \eqref{generating} does not converge down to zero transverse spin. We then point out a consequence of our results for the spectrum of certain Wilson lines in supersymmetric gauge theories. In subsection \ref{subsec:gravity}, we define a holographic defect in a three dimensional CFT and re-derive the large $s$ spectrum in eqs. (\ref{lightconetau}-\ref{cmin})  from a computation in the spirit of section~\ref{sec:ads}. In subsections \ref{subsec:renyi} and \ref{subsec:ising}, we turn to two examples in which the full defect OPE can be obtained by applying the inversion formula on a single block. 

Finally, in subsection \ref{subsec:onepoint} we point out an interesting common feature of all the examples.


\subsection{Defects in free theory}
\label{subsubsec:free}

As a first, simple, example we consider the defect spectra that can appear in the bulk-to-defect OPE of a free scalar ($\Delta_\phi=\tfrac{d-2}{2}$).
It was shown in section~B.1.1 of 
\cite{Billo:2016cpy} that only two towers of defect operators are allowed by the equations of motion:
\begin{align}
\ttr &= \Delta_\phi\,, \notag \\ 
\ttr &= \D_\phi+2-q-2s\,, \quad  s \leqslant\frac{4-q}{2}\,.
\label{eq:finitespinspectra}
\end{align}
The first set is the tower of transverse derivatives, which are not allowed to acquire anomalous dimension. This agrees with the lightcone analysis, and with the inversion formula \eqref{generating}, since all the operators in the bulk OPE of $\phi$ with itself have zero discontinuity, except the identity. In turn, as remarked in subsection \ref{eq:CH_singlebulk}, the inversion of the identity precisely yields the spectrum of the trivial defect, with the OPE coefficients \eqref{btrivial}. Those vanish for $m>0$ when $\D_\phi$ is at the unitarity bound, and indeed only the leading transverse twist trajectory appears in eq. \eqref{eq:finitespinspectra}.
What about the second tower in eq. \eqref{eq:finitespinspectra}? These are isolated operators at low spin, as enforced by the unitarity bound in eq. \eqref{eq:finitespinspectra}.
The lightcone expansion is blind to this kind of solutions. As we shall see now in a specific example, these operators also lie below the radius of convergence $s_\star$ of the inversion formula.

The simplest example of a non-trivial defect in free theory is obtained by integrating a free field on a dimension $p= \tfrac{d}{2}-1$ surface, which requires even $d \geqslant 4$ -- see \emph{e.g.} \cite{Billo:2016cpy} for more details. It follows from \eqref{eq:finitespinspectra} that the tower with bounded spin is only present if $d \leqslant 6$. In this case, a single defect operator with $s=0$ and $\Dhat=0$ is allowed -- the identity operator.
The two-point function of the free field is given by
\beq
\langle \phi(1,1) \phi(z,\bz) \rangle = \frac{1}{((1-z)(1- \bz))^{\Delta_\phi}}+ \frac{a_\phi^2}{(z \bz)^{\Delta_\phi/2}}\,,
\label{eq:WLcorr}
\eeq
which indeed differs from that of a trivial defect (first addend in \eqref{eq:WLcorr}) by the appearance of the defect identity (second addend in \eqref{eq:WLcorr}). 
We now want to use the Lorentzian inversion formula \eqref{generating} to recover the spectrum. We should check the behavior of $g(r,w)$
for $w \to0$ (or similarly $w \to \infty$) before dropping the arcs near $w=0$ and $w = \infty$ when going from \eqref{euclideaninv} to \eqref{eq:inversionformula}. 
The two-point function has the following asymptotics:
\beq
g(r, w)= r^{\Delta_\phi} \left\langle \phi(1,1) \phi\left(r w,\frac{r}{w}\right) \right\rangle \sim a_\phi^2\, w^0 +\OO( w^{\Delta_\phi} )\,, \quad \mathrm{for} \; w \to 0\,,
\eeq
and so from \eqref{sstar} we find that the inversion formula \eqref{generating} is valid for only for $s > s_\star=0$. Indeed, while the inversion of the first addend in \eqref{eq:WLcorr} reproduces the spectrum of the trivial defect, the second addend 
has zero discontinuity and does not contribute. Since the formula is not valid for $s=0$, this is not at odds with the presence of the identity in the defect OPE of $\phi$.

\subsection{Wilson lines in supersymmetric gauge theories}
\label{subsubsec:BPS}

It was conjectured in \cite{Lewkowycz:2013laa} that the one-point function of the stress tensor in the presence of certain BPS Wilson lines is related to the so-called Bremsstrahlung function as follows:\footnote{In \cite{Lewkowycz:2013laa}, the coefficient of the one-point function of $T_{\mu\nu}$ is called $h$, with $h=-a_T/d$.}
\beq
a_T=-  \frac{\Gamma\left(\frac{d-1}{2}\right)}{\pi^{\frac{d-3}{2}}} \frac{d(d-2)}{d-1} B~.
\label{atBconj}
\eeq
The Bremsstrahlung function measures the energy emitted by an accelerated charged particle at small velocities. The class of superconformal gauge theories in which the conjecture holds has not been completely explored yet. The relation \eqref{atBconj} has been checked for $1/2$ BPS Wilson lines in $\mathcal{N}=4$ SYM by direct computation of the two sides \cite{Okuyama:2006jc,Gomis:2008qa,Correa:2012at}. Evidence has been put forward also for $1/2$ and  $1/6$ BPS Wilson lines in ABJM \cite{Lewkowycz:2013laa,Bianchi:2017svd,Bianchi:2017ozk}, and for $1/2$ BPS Wilson lines in four dimensional $\mathcal{N}=2$ theories \cite{Fiol:2015spa}. 

Our results add an entry to the list of observables related to the Bremsstrahlung function. Indeed, via eqs. (\ref{cmin}-\ref{bmin}) we see that $B$ controls one of the leading contributions to the anomalous dimensions and the OPE coefficients of the transverse derivative operators at large spin. More precisely, recall that the Ward identities fix the three-point function $\braket{\phi\phi T_{\mu\nu}}$:
\beq
c_{\phi\phi T} = - \frac{d\,\D_\phi}{(d-1) S_d}, \qquad  S_d = \frac{2\pi^{d/2}}{\Gamma\left(\frac{d}{2}\right)}\,,
\label{cphiphiT}
\eeq
so that eqs. (\ref{cmin}-\ref{bmin}) become:\footnote{Our convention for the central charge is the same as eq.~(4.2) in \cite{Dolan:2000ut}, \ie, the two-point function of $T_{\mu\nu}$ has coefficient $C_T/S_d^2$. Of course, the combination $c_{\phi\phi}{}^Ta_T$ is independent of this choice, but notice that $c_{\phi\phi}{}^T=\frac{S_d^2}{C_T} c_{\phi\phi T}$.
\label{foot:CT}}
\begin{align}
c_\textup{min,T}&=-\frac{32\, \pi^{\frac{3}{2}}}{(d-1)^3} \frac{\Gamma(d+1)\Gamma\left(\frac{d+3}{2}\right) \Gamma(\D_\phi+1)}{\Gamma\left(\frac{d}{2}\right)^2\Gamma\left(\frac{d-2}{2}\right)
\Gamma\left(\D_\phi-\frac{d-2}{2}\right)}\frac{B}{C_T}~,
\label{cminTWL} \\  
b_\textup{min,T} &= c_\textup{min} \big(\gamma_E+\psi\left(d/2+1\right)\big)~.
\label{bminTWL}
\end{align}
Theories obeying the conjecture \eqref{atBconj} also have a protected scalar of dimension $d-2$ in their spectrum, belonging to the same superconformal multiplet as the stress tensor. The contribution of this scalar to the anomalous dimensions, and OPE coefficients, of transverse derivative operators is then of the same order as that of the stress tensor at large $s$, and we must take it into account. The coupling of this scalar and of the stress tensor, both to the Wilson line and to local operators that belong to certain short representations of the superconformal algebra, are related by the superconformal algebra. In these cases one can compute the correction to eqs.~(\ref{cminTWL}-\ref{bminTWL}) from the scalar operator. Depending on the theory one might need to consider other scalars of dimension $d-2$ (or lower).

Let us consider first the case of four-dimensional $\NN=4$ superconformal field theories in more detail. From the classification of irreducible highest-weight representations carried out in detail in \cite{Dolan:2002zh} (see also the summary in \cite{Beem:2016wfs}), we see that the only multiplet present in interacting theories that can contain twist-two conformal primaries in symmetric traceless representations is the stress tensor superconformal multiplet.\footnote{Concretely, apart from the stress tensor multiplet ($\BB^{\frac12,\frac12}_{[0,2,0]}$ in the notation of \cite{Dolan:2002zh}), the only multiplets that accommodate twist-two operators are the free vector multiplet ($\BB^{\frac12,\frac12}_{[0,1,0]}$), and multiplets containing conserved currents of spin greater than two ($\CC^{1,1}_{[0,0,0]}$ and $\BB^{\frac14,\frac14}_{[1,0,1],(0,0)}$), which should be absent in interacting theories \cite{Maldacena:2011jn,Alba:2013yda}.}
The twist-two symmetric traceless operators present in this superconformal multiplet are the stress tensor itself, the $SU(4)_R$ current -- which does not acquire a one-point function \cite{Billo:2016cpy}, and the superprimary -- a scalar of dimension two in the ${\bf 20}'$ representation of the R-symmetry group $SU(4)_R$.
From the two-point function of bulk operators neutral under $SU(2)_R$, we deduce that their defect OPE must contain transverse derivative operators obeying exactly (\ref{cminTWL}-\ref{bminTWL}), since the scalar ${\bf 20}'$ operators cannot be exchanged in the bulk OPE.
Considering the two-point functions of operators which transform non-trivially under $SU(4)_R$, we need to take into account the contribution of the dimension two scalar. In general, the relation between the coupling to the bulk operators of the scalar and the stress tensor is not known. However, if we consider the two point function of half-BPS bulk operators ($\BB^{\frac12,\frac12}_{[0,p,0]}$), the superconformal blocks obtained in \cite{Liendo:2016ymz} should allow for this relation to be determined and for a prediction to be made. This would amount to the anomalous dimension of long defect operators ($L^\star$ in the notation of \cite{Liendo:2016ymz}) whose transverse twists approach $p+2m$ as $s \to \infty$, and transforming in all representations of the $SO(5)_R$ R-symmetry, preserved by the line defect, that appear in the decomposition of the bulk operator's $[0,p,0]$ irrep of $SU(4)_R$.

For four-dimensional $\NN=2$ superconformal field theories there are various multiplets that can accommodate twist two operators. In particular, the stress-tensor supermultiplet once again contains a dimension two scalar as its superconformal primary, now neutral under the R-symmetry, this means that its contribution to (\ref{cminTWL}-\ref{bminTWL}) should generically be included. The relation between the one-point function of the scalar and the stress tensor is obtained in \cite{Fiol:2015spa}, however, the relation between their couplings to bulk operators is only known when considering half-BPS bulk operators, and only up to a sign. It would be interesting to work this out and add this correction to (\ref{cminTWL}-\ref{bminTWL}).

Finally, for three-dimensional $\NN=6$ or $\NN=8$ superconformal theories, once again there is a scalar in the stress tensor multiplet of dimension $d-2$, which transforms non-trivially under the R-symmetry.\footnote{Other scalars of dimension less or equal to one appear in multiplets that are either free, contain conserved currents of spin greater than one, or enhance the supersymmetry in the $\NN=6$ case, and so we do not need to consider them.}
This means that for R-symmetry singlet bulk operators this primary cannot be exchanged and the result in (\ref{cminTWL}-\ref{bminTWL}) holds.

\subsection{A holographic line defect in pure gravity}
\label{subsec:gravity}

In this subsection we present a holographic example. We find a solution of Einstein's equations with cosmological constant that we conjecture dual to a line defect in three dimensions, similar to the defect constructed in \cite{Horowitz:2014gva} for Einstein-Maxwell theory. We then compute anomalous dimensions of defect operators due to the exchange of the stress tensor. The computation is a holographic realization of the argument in section \ref{sec:ads}, and the result is in agreement with the lightcone bootstrap. This confirms the consistency of the whole picture.

\subsubsection*{The metric}

\begin{figure}[tb!]
    \centering
    \begin{subfigure}[t]{0.5\textwidth}
        \centering
\scalebox{0.6} 
{
\begin{pspicture}(0,-5.6966095)(11.974281,2.700828)
\psline[linewidth=0.048cm,linestyle=dashed,dash=0.16cm 0.16cm,arrowsize=0.05291667cm 2.0,arrowlength=2.0,arrowinset=0.4]{->}(3.765219,2.0973904)(3.765219,-5.6726093)
\rput(7.5097504,2.4373906){\Large $\xi=\infty$}
\rput(10.48975,2.4973905){\Large $\partial AdS_4$}
\psline[linewidth=0.048cm](3.765219,2.0973904)(8.765219,-2.9026096)
\psline[linewidth=0.048cm,arrowsize=0.05291667cm 2.0,arrowlength=2.0,arrowinset=0.4]{->}(3.765219,2.0973904)(11.55022,2.0973904)
\rput(9.74975,-2.8826096){\Large $\eta$}
\rput{-135.0}(5.871953,5.8870234){\psarc[linewidth=0.048](4.155219,1.7273904){1.86}{65.170654}{145.0}}
\rput{-135.0}(7.408091,4.953749){\psarc[linewidth=0.048](4.7300005,0.94260883){3.3446155}{76.338036}{153.9868}}
\rput{-135.0}(7.408091,4.9537497){\psarc[linewidth=0.048,arrowsize=0.033cm 5.0,arrowlength=2.0,arrowinset=0.4]{->}(4.7300005,0.94260895){3.3446155}{28.499008}{80.91735}}
\rput{-135.0}(5.871953,5.8870234){\psarc[linewidth=0.048,arrowsize=0.033cm 5.0,arrowlength=2.0,arrowinset=0.4]{->}(4.155219,1.7273904){1.86}{32.98852}{72.69947}}
\rput(6.26975,-1.6026095){\Large $\xi$}
\rput{90.0}(1.9774534,-4.9820476){\rput(3.4497504,-1.5026095){\Large $\xi=0$}}
\psline[linewidth=0.048cm,arrowsize=0.033cm 5.0,arrowlength=2.0,arrowinset=0.4]{<-}(8.645219,-2.7826095)(10.645219,-4.7826095)
\rput(3.5797503,2.3973904){\Large $\eta=\infty$}
\psdots[dotsize=0.4,linecolor=red,dotstyle=otimes](3.775219,2.0673904)
\end{pspicture} 
}
\caption{$(\xi,\eta)$}
    \end{subfigure}%
    ~ 
    \begin{subfigure}[t]{0.5\textwidth}
        \centering
        \scalebox{0.6} 
{
\begin{pspicture}(0,-3.21)(6.42,3.21)
\pscircle[linewidth=0.04,dimen=outer](3.21,0.0){3.21}
\psline[linewidth=0.04cm,linestyle=dashed,dash=0.16cm 0.16cm](3.21,3.17)(3.21,-3.17)
\psline[linewidth=0.04cm,arrowsize=0.05291667cm 2.0,arrowlength=1.4,arrowinset=0.4]{->}(3.228629,-0.05137085)(5.4913707,2.211371)
\psarc[linewidth=0.04,arrowsize=0.05291667cm 2.0,arrowlength=1.4,arrowinset=0.4]{<-}(3.63,1.12){0.79}{3.5035317}{117.03086}
\rput(4.393906,2.28){\Large $\theta$}
\rput(4.277656,0.38){\Large $r$}
\psdots[dotsize=0.4,linecolor=red,dotstyle=otimes](3.22,3.18)
\psdots[dotsize=0.4,linecolor=red,dotstyle=otimes](3.18,-3.18)
\end{pspicture} 
}
\caption{$(r,\theta)$}
    \end{subfigure}
\caption{(a) Depiction of the geometry in the coordinates of eqs.~\eqref{eq:metric} and~\eqref{eq:adsfactormetric}, at fixed $(\tau,\varphi)$. The metric has a conical singularity at $\xi=0$ (dashed line), and is asymptotically $AdS_4$ at large $\xi$. 
(b) Depiction of the geometry in global coordinates at fixed $(t,\varphi)$, see eqs.~\eqref{defect_metric_sph} and \eqref{metricglobal}. The conical singularity now extends along the diameter of the sphere (dashed line).}
\label{fig:coordsads}
\end{figure}
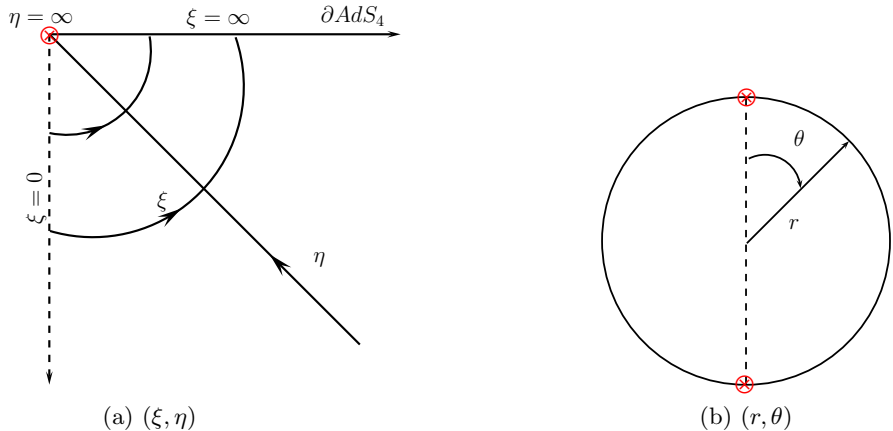
In view of the spirit of the argument given in section \ref{sec:ads}, in this subsection we work in Lorentzian signature with a time-like defect.
It will be convenient to write the metric using an $AdS_2-$slicing -- see for example \cite{Horowitz:2014gva}:
\beq
ds^2 = g(\xi) ds^2_{AdS_2} + \frac{d\xi^2}{f(\xi)} + \xi^2 d\varphi^2\,,
\label{eq:metric}
\eeq
where the $AdS_2$ factor is given by
\beq 
ds^2_{AdS_2}= -\eta^2 d\tau^2 + \frac{d\eta^2}{\eta^2}\, ,
\label{eq:adsfactormetric}
\eeq
and $\xi$ and $\varphi$ are radial and angular coordinates respectively. This metric has a manifest $SO(2,1) \times SO(2)$ symmetry, which is the expected symmetry group for a line defect in three dimensions. Notice that we used reparametrization invariance to fix the coefficient of $d\varphi^2$.

We would like this space to be asymptotically $AdS_4$, so we impose appropriate boundary conditions at large $\xi$:
\beq
f(\xi)  = \xi^2+1 + \OO\left(\frac{1}{\xi}\right)\,, \qquad
g(\xi)  = \xi^2+1 + \OO\left(\frac{1}{\xi}\right)\,.
\label{fginfty}
\eeq
The boundary can also be reached sending $\eta\to \infty$, which lands us on a one-dimensional subspace parametrized by $\tau$, which is the defect locus in the CFT -- see fig.~\ref{fig:coordsads}.
The boundary conditions in eq.~\eqref{fginfty} are enough to single-out a one-parameter solution to Einstein's equations. The function $f$ is fixed in terms of $g$:
\beq 
f(\xi) = \frac{4\xi g(\xi)(-1+3g(\xi))}{g'(\xi)(4g(\xi)+\xi g'(\xi))}\,.
\eeq
The function $g$ reads
\begin{align}
g(\xi) & = \xi^2+1+\frac{1}{3}(\xi^2+1)\left( -1 +\cos{\Theta(\xi)}-\sqrt{3} \sin{\Theta(\xi)} \right)\,,
\end{align}
where
\beq 
\Theta(\xi) = \frac{2}{3}\arcsin{\left(\frac{\Cm}{(1+\xi^2)^{\tfrac{3}{2}}}\right)}\, .
\label{thetacm}
\eeq
The parameter $\Cm$ controls the metric close to $\xi=0$. In particular, turning on a small $\Cm$ one gets
\beq
f(\xi) = 1 - \frac{4\,\Cm}{3\sqrt{3}} + \OO\left(\xi^2,\Cm^2\right)\,.
\eeq
This means that the metric develops a conical singularity at $\xi=0$, which breaks the isometries of $AdS_4$ down to the defect conformal group. The singularity is a conical defect for $\Cm>0$ and a conical excess in the opposite case. 

\subsubsection*{Single particle states and anomalous dimensions}

A basic entry in the $AdS$/CFT dictionary states that the energy of a state in $AdS$ with respect to global time equals the scaling dimension of the corresponding CFT operator. The discussion in section \ref{sec:ads} suggests that a large transverse spin operator is dual to a state describing a single particle with a large momentum in the $\varphi$ direction. Starting with $\Cm=0$, one obtains free propagation in empty $AdS$, and the spectrum of a trivial defect in the dual CFT. The conical singularity perturbs the gravitational potential, therefore the corresponding energy shift is dual to the contribution of the stress tensor to the scaling dimension of the CFT operator. In this subsection we demonstrate this matching. The computation is analogous to the determination of the anomalous dimension of large spin double-trace operators, as performed in section 2 of \cite{Fitzpatrick:2014vua}, which we follow closely -- see also \cite{Kaviraj:2015xsa}.

The first step is to switch to spherical coordinates:
\beq
\begin{split}
\tau & = \frac{1}{\cos{\theta} \sin{\rho(r)} \sec{t}-\tan{t}}\,,
\\
\eta & = \frac{2 (\cos{\theta} \sin{\rho(r)}-\sin{\tau})}{\sqrt{-2 \cos{2 \theta} \sin ^2{\rho(r)}+\cos{2
   \rho(r)}+3}}\,,
\\
\xi & = \sin{\theta} \tan{\rho(r)}\,,
\end{split}
\label{eq:sph_coords}
\eeq
where $\rho(r)=\arctan{r}$. In the new coordinates, the $AdS_2$ locus of the conical singularity extends along $t$ and along a diameter of the sphere at fixed $t$, see fig.~\ref{fig:coordsads}. It pierces the cylindrical boundary at the location of the defect -- at $\theta=0$ and $\theta=\pi$ -- as depicted in fig.~\ref{fig:aldaym}.\footnote{But notice that the labeling of coordinates is different from section \ref{sec:ads}, where holography was not involved. Also, the $AdS_2\times S^1$ frame induced on the boundary from eq. \eqref{eq:metric} is not the same as the $AdS_3\times S^1$ frame employed in section~\ref{sec:ads}.} The wave function of a particle that rotates fast in the $\varphi$ direction is peaked close to the boundary, therefore we only need to consider the large $r$ limit to obtain the leading order correction in $s$. As it can be seen from eq.~\eqref{thetacm}, this is equivalent to keeping only the leading order in $\Cm$ in all formulae. In this limit the metric is as follows:
\beq
\label{defect_metric_sph}
ds^2 = g_{tt}dt^2 + g_{rr}dr^2+g_{\theta \theta}d\theta^2 + g_{\varphi \varphi}d\varphi^2\,,
\eeq
where 
\begin{align}
\begin{split}
g_{tt} & = -r^2-1-\mathcal{C} \frac{4\left(r^2+1\right)}{3 \left(r^2 \cos{2 \theta}-r^2-2\right) \sqrt{3 r^2 \sin^2{\theta}+3} }\,,
\\
g_{rr} & = \frac{1}{(1+r^2)^2}\left(r^2+1- \mathcal{C}  \sqrt{\frac{2}{3}} \frac{4\left(r^2+1\right) \left(2 r^2\cos{2 \theta}+3 \cos{2 \theta}-2 r^2-1\right)}{3 \left(-r^2 \cos{2 \theta}+r^2+2\right)^{5/2}}\right)\,,
\\
g_{\theta \theta} & = r^2 + \mathcal{C} \sqrt{\frac{2}{3}} \frac{4r^2 \left(r^2\cos{2 \theta}+3 \cos{2 \theta}-r^2+1\right)}{3 \left(-r^2 \cos{2 \theta}+r^2+2\right)^{5/2}}\,,
\\
 g_{\varphi \varphi} & = r^2 \sin{\theta}\,.
\end{split}
\label{metricglobal}
\end{align}

The rest of the exercise is first-order perturbation theory applied to the Hamiltonian of a free scalar coupled to gravity. The unperturbed wave function is the following (see \eg, \cite{Penedones:2016voo})
\beq 
\psi_{m,\ell,s}(t,\rho,\theta,\varphi) = \frac{1}{N_{\Delta_{\phi} m \ell}}e^{-i E_{m,\ell}\, t}  Y^{s}_{\ell}(\theta, \varphi)  \sin^{\ell}{\rho}\, \cos^{\Delta_{\phi}}\!{\rho}\,
{_2}F_1(-m,\Delta_{\phi}+m+\ell,\ell + \frac{3}{2};\sin^{2}{\rho})\,,
\label{wavefunction}
\eeq
where
\beq
E_{m,\ell}=\Delta_\phi+2m+\ell\,,
\label{spenergy}
\eeq
$Y^s_\ell$ is a spherical harmonic and the normalization is
\beq 
N_{\Delta_{\phi} m \ell} = (-1)^m \sqrt{\frac{m! \Gamma \left(\frac{d}{2}+\ell \right)^2 \Gamma \left(\frac{2-d}{2}+m+\Delta_{\phi} \right)}{\Gamma \left(\frac{d}{2}+m+\ell \right) \Gamma (m+\ell +\Delta_{\phi} )}}\,.
\eeq
In empty $AdS_4$, $\psi$ is a descendant of a scalar primary of dimension $\Delta_\phi$. The dual interpretation of the labels is as follows: $m$ counts the number of Laplacians, while $\ell$ and $s$ are the angular momentum and its $\varphi$-component respectively. A feature of the spectrum of a trivial defect -- see eq.~\eqref{trivialspectrum} -- is that primary states of the defect correspond to bulk descendants with $s=\ell$, \emph{i.e.} the transverse spin equals the total spin of the bulk descendant. The quantity $\ell-|s|$ counts the number of derivatives along the defect. To isolate a defect primary, therefore, we choose $\ell=s$. Furthermore, we only consider the $m=0$ case for simplicity, that is, we focus on the leading transverse twist trajectory. It is then easy to see that $\psi_{0,s,s}(t,\rho,\theta,\varphi)$ in eq.~\eqref{wavefunction} is the wave function of a defect primary. Indeed, 
\begin{equation}
K_a (P_z)^s \ket{\phi}=0,
\end{equation} 
where $z$ is a complex coordinate parameterizing a plane and $a$ is a direction orthogonal to the same plane. Alternatively, one can check that the special conformal generator along the defect annihilates $\psi_{0,s,s}(t,\rho,\theta,\varphi)$.

The first order correction to the energies \eqref{spenergy} is then
\beq
\label{eq:deltaE}
\delta E = \langle m, s | \delta \Hm |m, s \rangle\,,
\eeq
where $\delta \Hm$ is the $\mathcal{O}(\Cm)$ term in the Hamiltonian of the scalar field. The computation is simplified by the fact that $\delta\sqrt{-g}=\mathcal{O}(\Cm^2)$. 
From the metric in \eqref{defect_metric_sph} the kinetic energy of all the coordinates except $\varphi$ receive corrections. 
We end up with three integrals
\beq
\label{eq:deltaEint} 
\delta E =  2 \Cm \int dr d\Omega_2 r^2 \left(\Im_1 + \Im_2 + \Im_3 \right)\,,
\eeq
where the integrands $\Im_i$ are given  by
\begin{align}
\Im_1 & =  \frac{2 r^{2 s } (\Delta +s )^2 \left(r^2+1\right)^{-\Delta -s -1} |Y_s^s(\theta ,\varphi )|^2}{3 \sqrt{3 r^2 \sin^2{\theta}+3} \left(r^2 \cos{2 \theta}-r^2-2\right)}\,,
\\
\Im_2 & = \frac{2 \sqrt{\frac{2}{3}} r^{2 s -2} \left(\left(2 r^2+3\right) \cos{2 \theta}-2 r^2-1\right) \left(r^2+1\right)^{-\Delta -s -1} \left(s -\Delta  r^2\right)^2
   |Y_s^s(\theta ,\varphi )|^2}{3 \left(-r^2 \cos{2 \theta}+r^2+2\right)^{5/2}}\,,
\\
\Im_3 & = -\frac{2 \sqrt{\frac{2}{3}} r^{2 s -2} \left(\left(r^2+3\right) \cos{2 \theta}-r^2+1\right) \left(r^2+1\right)^{-\Delta -s } |\partial_\theta Y^s_s(\theta ,\varphi )|^2}{3 \left(-r^2
   \cos{2 \theta}+r^2+2\right)^{5/2}}\,.
\end{align}
For $\ell=m=s$ the spherical harmonics simplify significantly:
\begin{align} 
|Y_s^s(\theta ,\varphi )|^2 & = \frac{1}{2 \pi^{\frac{3}{2}}}\frac{\Gamma(\frac{3}{2}+s)}{\Gamma(1+s)}(\sin{\theta})^{2s}\,,
\\
|\partial_\theta Y^s_s(\theta ,\varphi )|^2 & = \frac{1}{2 \pi^{\frac{3}{2}}}\frac{\Gamma(\frac{3}{2}+s)}{\Gamma(1+s)}s(\cos{\theta})^2(\sin{\theta})^{2s-2}\,.
\end{align}
The leading large $s$ asymptotics can be easily obtained using a saddle-point approximation. It turns out that of the three integrals in \eqref{eq:deltaEint} the first is the dominant one. After the dust has settled we find
\beq 
\delta E = -\Cm \frac{\Gamma (\Delta_{\phi} +1)}{3 \sqrt{3} \Gamma \left(\Delta_{\phi} -\frac{1}{2}\right)}\,s^{-\frac{1}{2}} \, .
\label{deltaEcomputed}
\eeq
The exponent of the correction matches the lightcone prediction eq.~\eqref{lightconeExponent}. In order to compare the prefactor with eq.~\eqref{cmin}, we need to compute the one-point function of the stress tensor in this geometry. This is an exercise in holographic renormalization \cite{deHaro:2000vlm}, which we perform in Euclidean signature. One first writes the metric in Fefferman-Graham coordinates \cite{Fefferman:2007rka}:
\beq
ds^2=\frac{1}{z^2}\left(dz^2+h_{\mu\nu}(x,z)dx^\mu dx^\nu\right).
\label{FefGra}
\eeq
We get
\begin{align}
h_{tt} &=1+\frac{z^2}{2}-\frac{2}{9\sqrt{3}}\Cm\frac{1}{\sin^3\theta}z^3+\mathcal{O}(z^4)\,, \\ 
h_{\theta\theta} &= 1-\frac{z^2}{2}-\frac{2}{9\sqrt{3}}\Cm \frac{1}{\sin^3\theta}z^3+\mathcal{O}(z^4)\,, \\ 
h_{\varphi\varphi} &=\sin^2\theta\left(1-\frac{1}{2} z^2+\frac{4}{9\sqrt{3}}\Cm \frac{1}{\sin^3\theta}z^3 +\mathcal{O}(z^4)\right).
\end{align}
The boundary now sits at $z=0$, and the expectation value of the stress-tensor is encoded in the coefficient of $z^d$ in the boundary expansion of $h_{\mu\nu}$:
\beq
\braket{T_{\mu\nu}}=\frac{d}{16\pi G_N} h^{(d)}_{\mu\nu}.
\eeq 
This immediately yields the one-point function on the cylinder. One can then apply a Weyl transformation to rewrite it on the plane. We report the result in Cartesian coordinates, which allows for an easy comparison with the form predicted by conformal invariance eq.~\eqref{onepointJ}. Recall that the defect extends along $t$, while we call $(x,y)$ the orthogonal coordinates:
\begin{align}
&\braket{T_{tt}}=-\frac{\Cm}{24 \sqrt{3} \pi G_N}\frac{1}{\abs{x_\perp}^3}\,,
\quad
\braket{T_{xx}}=\frac{\Cm}{24 \sqrt{3} \pi G_N}\frac{2y^2-x^2}{\abs{x_\perp}^5}\,, \\
&\braket{T_{yy}}=\frac{\Cm}{24 \sqrt{3} \pi G_N}\frac{2 x^2-y^2}{\abs{x_\perp}^5}\,,
\quad
\braket{T_{xy}}=-\frac{\Cm}{8 \sqrt{3} \pi G_N}\frac{xy}{\abs{x_\perp}^5}\,.
\end{align}
This allows to extract the coefficient of the one-point function:
\beq
a_T=-\frac{\Cm}{8\sqrt{3}\pi G_N}=-\frac{ C_T}{384\sqrt{3}}\,\Cm\,.
\label{aTgravity}
\eeq
The Newton constant is related to the central charge as follows:\footnote{Our conventions are summarized in footnote \ref{foot:CT}.}
\beq
G_N= \frac{\pi^{\frac{d}{2}-1} \Gamma(d+2)}{2 (d-1) \Gamma\left(\frac{d}{2}\right)^3 C_T}\,.
\eeq
The last ingredient is the three-point function coefficient $c_{\phi\phi T}$, which was given in eq. \eqref{cphiphiT}.
By plugging eqs.~\eqref{aTgravity} and \eqref{cphiphiT} in the lightcone bootstrap formula eq.~\eqref{cmin}, we precisely recover the anomalous dimension eq.~\eqref{deltaEcomputed}.

\subsection{The replica twist defect}
\label{subsec:renyi}

The replica defect arises in the computation of the geometric R\'{e}nyi entropies in QFT \cite{Calabrese:2004eu,Hung:2014npa,Bianchi:2015liz}. It is built as a boundary condition for the tensor product of $n$ copies of the QFT of interest -- $(\textup{QFT})^n$. In particular, the fields of consecutive copies are identified on a codimension one surface ending on the location of the defect, which has therefore codimension two. Given the replica defect supported on a surface $\Sigma$ lying on a constant time slice, consider the path-integral $Z_n(\Sigma)$ of the theory in the presence of the defect. Then the following holds:
\begin{align}
S_n(\Sigma)= \frac{1}{1-n}  \log \frac{Z_n(\Sigma)}{Z^n}, \qquad S_\textup{EE}(\Sigma)=\lim_{n\to 1}S_n(\Sigma)~,
\label{RenyiEE}
\end{align}
where $Z$ is the path-integral over a single copy of the QFT, and $S_n(\Sigma)$ and $S_\textup{EE}(\Sigma)$ are respectively the R\'{e}nyi and entanglement entropies of the density matrix obtained by tracing over the degrees of freedom enclosed in the region $\Sigma$. The second equation in \eqref{RenyiEE}, in particular, requires a definition of the theory that holds for complex values of $n$.
It was argued in \cite{Faulkner:2014jva} -- see also \cite{Balakrishnan:2017bjg} --  that an analytic continuation exists such that, when the original theory is a CFT, correlation functions preserve the features of a defect CFT, in particular the existence of a well defined defect OPE limit. The crucial step is to consider the orbifolded theory $(\textup{CFT})^n/\mathbb{Z}_n$, where only the local operators invariant under cyclic permutations of the copies are retained. In particular, the orbifold contains a unique stress tensor out of the $n$ coming from each copy.
This property is crucial in the proof of the quantum null energy condition \cite{Balakrishnan:2017bjg}. Notice also that the replica defect is trivial at $n=1$. Hence, it provides the example of a defect CFT with a perturbative parameter, such that the leading order deviation from the trivial defect is physically interesting in its own right.
 
 In this section, we apply the inversion formula to a scalar two-point function and obtain the anomalous dimensions of a class of defect operators at leading order in $(n-1)$. The result matches the computation in \cite{Balakrishnan:2017bjg}, performed by direct OPE decomposition of the correlator.
The external operator is the $\mathbb{Z}_n$-symmetrization of a scalar primary $\phi$ belonging to a single copy:
\beq
[\phi] = \sum_{k=1}^n \phi^{(k)}~,
\label{orbiO}
\eeq
where the bracket denotes orbifold operators, and $\phi^{(k)}$ stands for the operator $\phi$ inserted in the $k$-th copy. We call $[\phi]$ a single-copy orbifold operator, while multiple-copy operators are the $\mathbb{Z}_n$-symmetrization of products of operators on different replicas (\emph{e.g.} $\phi_1^{(i)}\phi_2^{(j)}$ with $i\neq j$). The bulk OPE $[\phi]\times [\phi]$ contains single-copy and double-copy operators. The former correspond to the primaries in the fusion $\phi\times \phi$, while the latter arise from the non-singular fusion $\phi^{(i)}\times \phi^{(j)}$, $i\neq j$, and have twist $\tau=2\Delta_\phi+2m$ with integer $m$. This fusion rule is independent of the number of copies, and survives in the analytically continued theory. From the explicit analytic continuation of single-copy operators \cite{Balakrishnan:2017bjg}, one sees that the only single-copy operator with a vev of order $(n-1)$ is the stress-tensor. However, a single block is not crossing symmetric, and the double-copy operators come to the rescue and also acquire a one-point function.
Since the twists of the latter differ from $2\D_\phi$ by an even integer, they do not contribute a discontinuity to the two-point function $\braket{[\phi][\phi]}$, and so do not affect the OPE coefficients nor the anomalous dimensions of defect operators. Hence, the defect spectrum is simply found by inverting the stress tensor block. The one-point function of the stress-tensor at order $(n-1)$ is known exactly \cite{Hung:2011nu,Perlmutter:2013gua,Hung:2014npa}:\footnote{Our convention is related to the one in \cite{Hung:2014npa} as follows: $a_T=-d h_n/(2\pi)$. See also footnote \ref{foot:CT}.}
\beq
\left.\partial_n a_T \right|_{n=1} = -d\,\pi^{d/2} \frac{\Gamma\left(d/2\right)}{\Gamma\left(d+2\right)}\frac{C_T}{S_d^2}~,\qquad S_d = \frac{2\pi^{d/2}}{\Gamma\left(\frac{d}{2}\right)}\,,
\eeq
and the same is true for the OPE coefficient, see eq.~\eqref{cphiphiT}. To obtain the leading transverse-twist primaries, we only need to evaluate eq.~\eqref{gammaStress} for $q=2$, and the result is
\beq
\ttr= \Delta_\p - (n-1)\frac{\D_\p}{ d-1} \frac{\left(\D_\p-\frac{d}{2}+1\right)_s}{\left(\D_\p\right)_s}+\OO\big((n-1)^2\big)~,
\label{replicaSpectrum}
\eeq
where $(a)_s=\Gamma(a+s)/\Gamma(s)$.
This is the correct result \cite{Balakrishnan:2017bjg}.\footnote{In fact, our anomalous dimension is twice the one in eq.~(3.25) of \cite{Balakrishnan:2017bjg}. However it matches their eq.~(B4), so we attribute the discrepancy to a typo.} However, to confirm its validity down to finite transverse spin we need to know the behavior of the two-point function $\braket{[\phi][\phi]}$ as $w\to 0$. In the absence of a general bound, we resort to the explicit form computed in \cite{Balakrishnan:2017bjg}:
\begin{multline}
g(z,\bz) = \left(\frac{(1-z)(1-\bz)}{(z\bz)^{1/2}}\right)^{-\Delta_\phi} \left(1+ c_{\phi \phi T}\,a_{T}\, f_{d,2}(z, \bz) +I+ \OO(n-1)^2 \right)\,, \\
I=- (n-1) \int_0^{-\infty} \frac{d\lambda}{(\lambda-1)^2} 
\left(1+\frac{1}{w \xi} (1-\lambda)+\frac{w}{\xi}\left(1-\frac{1}{\lambda}\right)\right)^{-\D_\phi}\,,
\end{multline}
where $f_{d,2}(z, \bz)$ is the bulk stress-tensor block and $\xi$ is defined in eq.~\eqref{xicross}. The integral $I$ can be seen in a bulk block decomposition to encode the contributions of the aforementioned exact double-twist operators.
Expanding $I$ at leading order in $w$ one finds
\begin{multline}
\left(\frac{(1-z)(1-\bz)}{(z\bz)^{1/2}}\right)^{-\Delta_\phi} I =(n-1)  w^\D\, \pi \D \csc \pi \D\,\big(1+O(w)\big) \\
+(n-1) w\, \D r^{1-\D} \frac{\Gamma(\D) \Gamma(\D-1)}{\Gamma (2\D)}
{}_2F_1\left(\D-1,\D,2\D,1-\frac{1}{r^2}\right)\,\big(1+O(w)\big)\,.
\end{multline}
This asymptotics, together with the single block asymptotics eq.~\eqref{smallwblock}, ensure that $s_\star<0$ -- see eq.~\eqref{sstar} -- and so eq.~\eqref{replicaSpectrum} holds for all $s$.

\subsection{The Ising twist defect}
\label{subsec:ising}

There exists a conformal defect with codimension two in the 3d Ising model, as supported by numerical evidence \cite{Billo:2013jda} and also by results from the epsilon expansion and the conformal bootstrap \cite{Gaiotto:2013nva}. Local operators odd (even) under the $\mathbb{Z}_{2}$ flavor symmetry of the 3d Ising CFT are multi-valued (single-valued) around the twist defect. As a consequence, the $\mathbb{Z}_{2}$ odd (even) defect spectrum take half-integer  (integer) values of the transverse spin $s$.

Following the literature, let us dub $\psi_{s}$ the leading transverse twist primaries in the defect OPE of the spin field $\sigma$:
\beq
\si \sim \sum_s \psi_s + \textup{higher } \ttr~,\qquad s\in \mathbb{N}+\frac{1}{2}~.
\eeq 
The dimensions and OPE coefficients of the $\psi_{s}$ have been calculated in the epsilon expansion in \cite{Gaiotto:2013nva}, \emph{i.e.} by setting $d=4-\epsilon$ and keeping $q=2$. To leading order in $\epsilon$,
\begin{eqnarray}
\ttr_{\psi_{s}}&=& 1-\left(\frac{1}{2}+\frac{1}{24s}\right)\epsilon +O(\epsilon^{2})~,          \label{psi} \\
|b_{\sigma \psi_{s}}|&=& 1+\frac{\psi(1)-\psi(s+1)}{4}\epsilon +O(\epsilon^{2})   ~, \qquad \psi(z)=\frac{d}{dz} \ln{\Gamma(z)}~.         \label{psicoeff}
\end{eqnarray} 
Let us interpret these values from the lightcone bootstrap point of view. The scaling dimension of $\sigma$ is
\beq
\D_\sigma=1-\frac{\epsilon}{2}+O(\epsilon^2)~,
\label{deltasigma}
\eeq
so the $\psi_s$ are easily identified as the leading trajectory of transverse derivative operators. The fusion $\sigma\times \sigma$, at the leading non-trivial order in $\epsilon$, can be written as follows:
\beq
\sigma\times \sigma \sim 1+\varepsilon+\{\tau=2\D_\sigma\}+\textup{higher twists}+O(\epsilon^2)~,
\eeq
where $\varepsilon$ is the energy operator and $\{\tau=2\D_\sigma\}$ denotes the conserved currents of the free theory, which do not acquire anomalous dimension at this order \cite{Wilson:1973jj} -- see also \cite{Alday:2015ota} and \cite{Gliozzi:2017gzh} for a more general understanding of this fact. The higher-twist primaries are decoupled in the free theory, and so their OPE coefficient is $O(\epsilon)$. Hence, they also appear as operators with $\tau=2\D_\sigma+2m$ in this OPE. All together, the only primary contributing a discontinuity is $\varepsilon$, which is therefore fully responsible for eqs. \eqref{psi} and \eqref{psicoeff}. The required OPE data were presented in \cite{Gaiotto:2013nva}:
\begin{eqnarray}
\Delta_{\varepsilon}=2-\frac{2\epsilon}{3}+O(\epsilon^2)~, \qquad c_{\si\si\varepsilon}a_{\varepsilon}=-\frac{1}{8}+O(\epsilon)~.
\label{epsilondata}
\end{eqnarray}    
In fact, the full result \eqref{psi}, \eqref{psicoeff} is encoded in the leading transverse spin correction. Indeed, plugging eq.~\eqref{epsilondata} in eq.~\eqref{cmin}, we reproduce the value of the anomalous dimension $\gamma_{s,0}=-\frac{1}{24s}\epsilon$. Furthermore, the correction $b_\textup{min}$ in eq.~\eqref{bmin} is $O(\epsilon^2)$, and indeed the square root of eq.~\eqref{btrivial} reduces to \eqref{psicoeff}. Despite the simplicity of the result, it is not obvious why the large $s$ expansion of the anomalous dimension should truncate at order $1/s$. We can address the question by means of the inversion formula. By evaluating the single block contribution eq.~\eqref{gammaScalar} with $q=2$, $d=4-\epsilon$ and the CFT data in eqs.~\eqref{deltasigma} and \eqref{epsilondata}, we indeed get, for $s>0$,
\beq
\gamma_{s,0}=-\frac{\epsilon}{24}\frac{1}{s+1}\, {}_2 F_1 (1,1,s+2,1)+O(\epsilon^2)=
-\frac{\epsilon}{24s }+O(\epsilon^2)~.
\label{gammapsi}
\eeq
It is interesting to notice that each bulk collinear block contributes an infinite series in $1/s$, and the final result emerges from infinitely many exact cancellations.
As before, in principle we must check the small $w$ limit of the $\braket{\si\si}$ correlator to confirm that the leading-twist spectrum \eqref{psi}, as found with the inversion formula, is complete. We could not obtain this behavior analytically, however, a numerical analysis of the correlator of \cite{Gaiotto:2013nva}, suggests that for $w \to 0$ the $\langle \sigma \sigma \rangle$ correlator behaves like $\sim \alpha_r w^0$, where $\alpha_r$ is a coefficient that depends on the radius $r$. Thus $s_{\star} =0$ and the inversion formula \eqref{eq:inversionformula} only holds for $s>0$.
We were not able to find the contribution of a single scalar block to the OPE coefficient in closed form. However, from the computation of the OPE coefficient as an infinite sum, discussed in section \ref{eq:CH_singlebulk}, we can easily check that the same cancellations are in place: this time, after appropriately including the Jacobian in eq.~\eqref{eq:Jac}, no contribution is left at order $\epsilon$. Therefore, we also recover eq.~\eqref{psicoeff}. 
We can also predict the existence of the higher transverse twist primaries, with $\ttr \to \Delta_\sigma + 2m$ at large spin, whose OPE coefficients, for $m \neq 0 $, are of order $\epsilon$ as clear from the fact that \eqref{btrivial} vanishes for $m>0$ and $\D_\p$ at the unitarity bound.

Let us conclude with some comments on the $\mathbb{Z}_2$ even defect spectrum. In free theory, the leading transverse twist operators are bilinear of the $\psi_s$, and all operators $\psi_{s_{1}}\psi_{s_{2}}$ with integer transverse spin $s_{1}+s_{2}=s$ have the same transverse twist $\ttr=2-\epsilon$. This $\floor{\frac{s+1}{2}}$-fold degeneracy is lifted at the Wilson-Fisher fixed point, and we parametrize the eigenvalues of the matrix of anomalous dimensions as follows: 
\beq
\ttr_{s,j}=2+\epsilon\left(\delta_{s,j}-1\right)  +O(\epsilon^2)     \,.    \label{WFdim}
\eeq
 In \cite{Gaiotto:2013nva}, it was pointed out that the following accumulation points exist at infinite transverse spin:
 \beq
 \delta_{\infty,j}=-\frac{1}{12(2j-1)}~,\qquad j=1,2,\dots\,.
 \label{accumnoder}
 \eeq
 The results of section \ref{sec:lightdefect} predict an additional accumulation point: the leading transverse derivative of the energy operator $\varepsilon$, that is,
  \beq
 \delta_{\infty,0}=\frac{1}{3}~.
 \label{accumder}
 \eeq
In fact, it is not hard to see that this accumulation point exists,\footnote{We thank D.~Maz\'{a}\v{c} for sharing with us an elegant analytic proof of this fact.} and that furthermore eqs. \eqref{accumnoder} and \eqref{accumder} comprise all the anomalous dimensions of this class of operators. As for the accumulation points \eqref{accumnoder}, those are not transverse derivatives of $\varepsilon$, and therefore we should expect their OPE coefficient to be subleading at large $s$. We did not check this fact. Both the lightcone bootstrap and the use of the inversion formula are complicated by the presence of infinitely many bulk blocks contributing already at order $\epsilon$, so we leave this analysis for future work.

\subsection{\texorpdfstring{On the sign of $\braket{T_{\mu\nu}}$}{On the sign of <Tmunu>}}
\label{subsec:onepoint}

In this subsection, we observe a feature of the one-point function of the stress tensor in the presence of a defect which is found in many examples. Let us start from the holographic defect discussed in subsection \ref{subsec:gravity}, and in particular from eq. \eqref{aTgravity}.
The attractive nature of gravity indicates that the parameter $\Cm$ should be positive -- see \eg, eq.~\eqref{deltaEcomputed}. From eq.~\eqref{aTgravity}, we deduce that $a_T<0$. The same happens in all unitary defect conformal field theories we are aware of: in particular, the Wilson lines in free theories discussed in \cite{Billo:2016cpy}, the supersymmetric Wilson line briefly discussed in subsection~\ref{subsubsec:BPS}, and the Renyi twist operator \cite{Hung:2014npa,Bianchi:2015liz}. Let us briefly explain why this is the case for the latter. In~\cite{Hung:2011nu}, eq.~(4.21), the following expression was derived:
\beq
a_T(n)= -\frac{d}{d-1}n\left(\mathcal{E}(T_0)-\mathcal{E}(T_0/n)\right)\,.
\eeq
Here, $n$ is the so-called replica index, $\mathcal{E}(T)$ is the energy density of a thermal state at temperature $T$ on the $(d-1)$-dimensional hyperbolic space $H^{d-1}$ of unit radius and $T_0=1/2\pi$. Stability of the canonical ensemble implies that the specific heat $\frac{d E}{dT}$ is positive. Here $E$ is the total energy, but since $\braket{T_{\mu\nu}}$ is constant on this background, the same property is valid for $\mathcal{E}(T)$.  We deduce that $a_T<0$ if $n>1$. On the other hand, exactly at $n=1$ the defect becomes trivial, and for $n<1$ the theory becomes non unitary: for instance, the two-point function of the displacement operator \cite{Billo:2016cpy} has negative norm $C_D<0$ for $n$ slightly smaller than 1, as it can be deduced by the following equation \cite{mark0,Bianchi:2015liz,Faulkner:2015csl}: 
\beq
\left.\partial_n C_D\right|_{n=1}=\frac{2\pi^2}{d+1}\frac{C_T}{S_d^2}\,.
\eeq
 
The evidence put forward so far makes it plausible that $a_T\leq 0$ in all unitary defect conformal field theories. It would be interesting to prove this statement or look for a counter-example, but we leave this for future work.

\section{Discussion and outlook}
\label{sec:discussion}

The main technical tool used in this paper is the inversion formula \eqref{eq:inversionformula}. The formula expresses the defect spectrum and the OPE coefficients as an integral over the discontinuity of the two-point function. The integral kernel is analytic in $s$, which means that the spectrum of every defect CFT is organized in Regge trajectories $\Dhat(s)$, at least down to some minimum spin $s_\star$. The value of $s_\star$ is finite as long as the correlator is polynomially bounded in a certain double lightcone limit -- see fig.~\ref{fig:rhobulk}.

At large $s$, the generating function \eqref{generating} localizes close to $\bz=1$, and is therefore dominated by the bulk-channel OPE limit of the discontinuity. This fact puts on solid ground the result first obtained in section \ref{sec:lightdefect} via the lightcone bootstrap: in every defect CFT, a subset of the trajectories asymptote the spectrum of the trivial defect at large $s$ -- see eq.~\eqref{deftwistlead}. We call transverse derivatives the operators on these trajectories. Since $s$ is a global charge for the defect spectrum, it is interesting to contrast this result with the large charge sector of an ordinary CFT \cite{Hellerman:2015nra,Monin:2016jmo,Jafferis:2017zna}. There, an operator with a large charge creates in radial quantization a state with homogeneous charge and energy density. The scaling of the low-lying spectrum  with the charge is not linear, and therefore incompatible with the presence of the transverse derivatives. In fact, as we saw in section \ref{sec:ads}, the semiclassical picture in this case is very different from a condensed matter phase. Since a defect CFT is non local, there is no notion of energy density on the defect, and once seen from the bulk, the transverse derivative breaks a spacetime rotational symmetry rather than a global one.

Finite $s$ corrections to the spectrum and the OPE coefficients can in principle be computed systematically, by applying the inversion formula block by block, as a convergent expansion.
In contrast to the lightcone bootstrap where only an asymptotic expansion is obtained, by using the inversion formula one gets a better and better approximation to the defect spectrum as more bulk blocks are added.
However, in practice we achieve analytic control only by replacing the bulk-channel blocks with their expansion in powers of $z$. Although this procedure is in general incorrect, it is sufficient to resum part of the large $s$ expansion of the anomalous dimensions of the transverse derivatives. In fact, if the discontinuity receives contributions only by a finite number of blocks, the procedure yields the full spectrum. This happens sometimes in a perturbative setup, as we saw in the examples and we further remark in the next subsection. 

The perturbative setting also highlights a difference between this inversion formula and the inversion formula for the four-point function. The discontinuity of a single logarithm, unlike the double discontinuity, does not vanish. This has an immediate consequence for defects embedded in theories with a slightly broken higher spin symmetry: the transverse derivatives acquire anomalous dimension at leading order in the breaking parameter. We saw this happening in subsection \ref{subsec:ising}. The opposite is true in the case of the four-point function, a fact recently exploited in \cite{Alday:2017zzv} to efficiently apply large-spin perturbation theory order by order in the $\epsilon$ expansion.

Some important aspects concerning the inversion formula remain to be clarified. Firstly, we did not place an upper bound on $s_\star$. Theories whose correlators are not polynomially bounded in the asymptotic region mentioned earlier may not display Regge trajectories. It would be very interesting to prove an upper bound, or to find a counter-example. A bound cannot be proven in the same way as in \cite{Caron-Huot:2017vep}, because the bulk channel OPE is not positive, a fact that is also tied to the appearance of the discontinuity in the formula, which has no definite sign, unlike the double discontinuity of the Caron-Huot formula.

Secondly, it would be important to set up a procedure that works order by order in the small $z$ expansion and at the same time does not spoil the convergence of the formula. This problem is related to the necessity of dealing with the towers of double- and multi-twist operators in the bulk channel OPE. In the case of the four-point function, this issue has been solved \cite{Simmons-Duffin:2016wlq,Caron-Huot:2017vep}.

\subsubsection*{Outlook}

The results of this work suggest various future directions of research. It would be interesting to study the large $s$ spectrum in other examples, possibly beyond leading order in perturbation theory, or rather in strongly coupled scenarios. For instance, in section \ref{sec:ads} we pointed out that the spectrum of a Wilson line at large $N$ supports operators whose transverse twist measures the energy of a heavy-light quark pair -- see eq. \eqref{singletracedefect}. It would be interesting to explore this case with the inversion formula, and to study the transverse derivatives as well. Notice that one-point functions of double trace operators are suppressed at large $N$, so the inversion formula suggests that the anomalous dimension of the transverse derivatives receives contribution only from the exchange of single trace operators at leading order in $1/N$.

From a technical point of view, the inversion formula \eqref{eq:inversionformula} can be obtained via manipulations that are much simpler than in the case of the four-point function. The formula itself depends on a kernel explicitly known in closed form. This makes the defect case a good playground, both for applications and for generalizations of the formula. In particular, deriving the inversion formula in the case of external spinning operators might be a doable task.
For this purpose, a method similar to the one presented in \cite{Karateev:2017jgd} may be useful.
Also in this last case, the presence of the transverse derivatives at large $s$ is expected from the lightcone bootstrap analysis, which is essentially unchanged. Their scaling dimensions should again approximate the trivial defect, \ie, 
\beq
\ttr \simeq \tau_\phi +2 m~,\qquad s\to\infty~.
\eeq

Finally, one obvious direction of investigation concerns the inversion of the bulk OPE, and it is work in progress. 
Under many respects, this problem is more similar to the one solved by the inversion formula for the four-point function \cite{Caron-Huot:2017vep}. Indeed, the inversion of the bulk OPE would yield the same spectrum, albeit  associated to different OPE coefficients.
This inversion formula would recover bulk information from the defect OPE in the limit when one of the bulk operators is light-like separated from the defect, \ie, the opposite limit to the one considered in \ref{sec:lightdefect}. A preliminary lightcone bootstrap analysis seems to show that the existence of double twist operators is sufficient to satisfy the crossing constraint in this limit at the leading order, so that no obstruction exists at this level for a CFT to support a conformal defect.
An inversion formula for the bulk OPE, combined with \eqref{eq:inversionformula}, could also allow for an iterative procedure to obtain approximate solutions to crossing symmetry starting from only a few operators, similarly to what was done in \cite{Simmons-Duffin:2016wlq} for the $3d$ Ising model.
Furthermore, the crossed-channel -- which in this case is the defect channel -- is positive, and this may provide a better control over the convergence of the formula. It would also be interesting to see if the positivity of the defect OPE can be used to answer the question raised in subsection \ref{subsec:onepoint} on the sign of the one-point function of the stress tensor.

\acknowledgments

We have greatly benefited from discussions with
L.~Bianchi,
D.~Maz\'{a}\v{c},
J.~Penedones,
B.~van~Rees,
S.~Rychkov,
V.~Schomerus,
E.~Trevisani,
S.~Giombi,
R.~Rattazzi.
M.M. is especially grateful to D.~Gaiotto, for crucial suggestions in the early stage of this project.
M.L., P.L. and M.M. thank the Galileo Galilei Institute for Theoretical Physics for hospitality and the INFN for partial support during the completion of this work during the workshop ``Conformal field theories and renormalization group flows in dimensions $d>2$''. M.L., P.L. and M.M. thank ICTP-SAIFR in S\~{a}o  Paulo for hospitality during the Bootstrap 2017 workshop, and the Simons Collaboration on the Non-perturbative Bootstrap for providing many stimulating workshops and conferences during which this work carried out. S.S. is thankful to DESY Hamburg and EPFL for hospitality and support for visits during the completion of this work. We also thank the participants of the Simons Collaboration's kick off, ``Bootstrap and Quantum Gravity'' and ``Numerical bootstrap'' meetings for discussions. M.L. was partially supported  by the German Research Foundation (DFG) via the Emmy Noether program ``Exact results in Gauge theories''.
M.M.  is supported by the Simons
Foundation grant 488649 (Simons collaboration on the non-perturbative bootstrap) and by
the National Centre of Competence in Research SwissMAP funded by the Swiss National Science
Foundation.


\appendix

\section{Conformal blocks}
\label{sec:blocks}

Here we review the kinematics of a two-point function of bulk operators in the presence of a flat defect of codimension $q$ in $d$ dimensions, following \cite{Billo:2016cpy}.
Recall that we separate the space time indices ($\m=0,\dots, d-1$) into two subsets, those orthogonal to the defect are labeled by $i=0,\dots, q-1$ and those parallel to the defect by $a=q,\dots, d-1$.
The the two-point function of bulk identical scalars depends on two conformally invariant cross-ratios, and we parametrize it by
\beq
\braket{\phi(x_1)\phi(x_2)}=\frac{1}{(\abs{\xper_1} \abs{\xper_2})^{\D_\phi}} g(\chi,\eta)\,,
\label{eq:2pointApp}
\eeq
where we use $\xper = \{x^i\}$ and  $\xpar= \{x^a\}$, and the cross-ratios are given by\footnote{Here $\eta$ is what was called $\cos \phi$ in \cite{Billo:2016cpy}.}
\beq
\chi =
 \frac{\abs{\xpar_{12}}^2+\abs{\xper_1}^2+\abs{\xper_2}^2}{\abs{\xper_1}\abs{\xper_2}}= \frac{1+z\bz}{(z\bz)^{1/2}}=
 \frac{1}{r}+r\,, \qquad
\quad \eta =
 \frac{x_{1i} x_2^i}{\abs{\xper_1}\abs{\xper_2}}=\frac{z+\bz}{2 (z\bz)^{1/2}}=\frac{1}{2}\left(w+\frac{1}{w}\right) \,.
 \label{crossratio}
\eeq
As discussed in section \ref{sec:lightdefect} we work in Lorentzian signature with a space-like defect, and with the geometry shown in fig.~\ref{diamond}, so for convenience above we have shown the relation between the cross-ratios $(\chi,\eta)$ and the lightcone $(z,\bz)$ coordinates, as well as the radial coordinates $r$ and $w$.
The crossing equation reads (or equivalently  \eqref{eq:crossingsy} in $(z,\bz)$ coordinates)
\beq
g(\chi,\eta) =\xi^{-\D_\phi}\sum_{O} c_{\phi\phi O}\, a_O \,f_{\Delta,J}(\chi,\eta) =\sum_{\widehat{O}} (b_{\phi\widehat{O}})^2 \,\widehat{f}_{\widehat{\tau},s}(\chi,\eta)\,,
\label{eq:crossingsyApp}
\eeq
where we defined the transverse twist $\widehat{\tau}=\widehat{\D}-s$, and 
\beq
\xi=\chi-2\eta=\frac{(1-z)(1-\bz)}{(z\bz)^{1/2}}\,.
\label{xicross}
\eeq
The defect conformal blocks are known exactly \cite{Billo:2016cpy}\footnote{The blocks here differ from the ones of  \cite{Billo:2016cpy} by a factor of $2^s$.}
\beq
\fy_{\widehat{\tau},s}(\chi,\eta) = \binom{s+\frac{q}{2}-2}{\frac{q}{2}-2}^{-1}\, C_s^{(q/2-1)}\left(\eta\right)
\, _2F_1\left(\frac{\widehat{\Delta} }{2}+\frac{1}{2},\frac{\widehat{\Delta} }{2};\widehat{\Delta} +1-\frac{p}{2};\frac{4}{\chi^2}\right)\,,
\label{eq:defblockApp}
\eeq
where $C_n^{(m)}\left(x\right)$ is a Gegenbauer polynomial, and changing variables to $(z,\bz)$ the above block can be re-written as \eqref{zdefblock}, after using the following relation
\beq
\binom{s+\frac{q}{2}-2}{\frac{q}{2}-2}^{-1} C_s^{q/2-1}\left(\frac{w}{2}+\frac{1}{2 w}\right)=w^{-s} {}_2F_1\left(-s,\frac{q}{2}-1,2-\frac{q}{2}-s,w^2\right)\,.
\label{gegenhyperApp}
\eeq
When $q$ is even, an order of limits ambiguity arises in the definition of the hypergeometric function in \eqref{gegenhyperApp}, with the equality holding if we first take $s$ to be integer, and then $q$ to be even. This prescription is always assumed throughout the paper.
The normalization is chosen such that, given a leading contribution to the defect OPE of the kind 
\beq
\phi(x^\m) \sim b_{\phi \wh{O}}\, \abs{\xper}^{-\D_\phi+\wh{\D}-s} x^{i_1} \cdots x^{i_s} \wh{O}_{i_1\dots i_s} (x^a)
+ \ldots\,,
\label{leaddope}
\eeq
the contribution of $\wh{O}$ to the two-point function is as in eq.~\eqref{eq:crossingsyApp}, provided that the two-point function of defect primaries is normalized as
\begin{equation}
\langle\widehat{O}^{i_1...i_s}(x_1^a)\widehat{O}^{j_1...j_s}(x_2^a)\rangle 
=2^{s} \frac{ \mathcal{P}^{i_1...i_s; j_1...j_s}}{(x_{12}^2)^{\widehat{\Delta}}}\,.
\label{2pdefectph}
\end{equation}
Here $\mathcal{P}$ is the projector onto symmetric and traceless tensors, which can be defined in terms of the Todorov operator (see \cite{Billo:2016cpy} for more details)
\begin{multline}\label{sttprojector}
\mathcal{P}^{i_1...i_s; j_1...j_s}\equiv \frac{1}{s!\left(\frac{q}{2}-1\right)_s}D_{i_1}\dots D_{i_s}w_{j_1}\dots w_{j_s}, \\
D_i =\bigg(\frac{q-2}{2}+w^j \frac{\partial}{\partial w^j}\bigg)\frac{\partial}{\partial w^i}-\frac{1}{2}w_i\frac{\partial^2}{\partial w^j \partial w_j}\,.
\end{multline}
Finally, we denote the defect OPE coefficient of the identity, $b_{\phi \hat{\mathbb{1}}}$, by $a_\phi$, such that the one-point function of a bulk primary of dimension $\Delta$ and spin $J$ is
\beq
\langle \phi(x^\mu , z^\mu ) \rangle = \frac{a_\phi}{|\xper|^{\Delta_\phi}} \left( \frac{(z^i x^i)^2}{|\xper|^2}-z^i z^i\right)^{\frac{J}{2}}\,,
\label{onepointJ}
\eeq
where $z^\mu$ is an auxiliary null vector used to contract the indices of the spin $J$ operator (see \cite{Billo:2016cpy} for how to recover the index structure encoded in the polynomial dependence in $z$) .

\section{Convenient coordinates for the bulk channel OPE}
\label{sec:rhobulk}

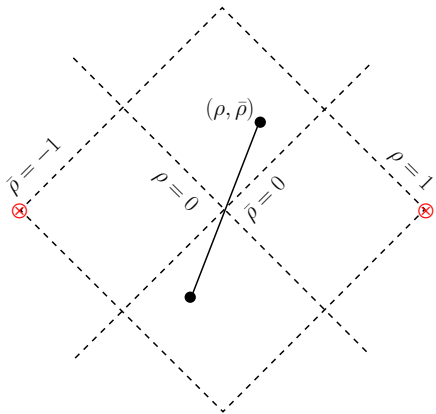
\begin{figure}[htb!]
\centering
\scalebox{0.5} 
{
\begin{pspicture}(0,-5.4234624)(11.22,5.4234624)
\psdots[dotsize=0.3](4.74,-2.3019192)
\psdots[dotsize=0.3](6.6,2.3580809)
\psline[linewidth=0.04cm](4.78,-2.2619193)(6.58,2.3780808)
\rput{-44.895756}(1.6327958,3.9520829){\psframe[linewidth=0.04,linestyle=dashed,dash=0.16cm 0.16cm,dimen=outer](9.400117,3.8408313)(1.7584544,-3.8008313)}
\psline[linewidth=0.04cm,linestyle=dashed,dash=0.16cm 0.16cm](1.6292253,4.0693536)(9.426561,-3.7503076)
\psline[linewidth=0.04cm,linestyle=dashed,dash=0.16cm 0.16cm](9.497724,3.878191)(1.6780626,-3.9191449)
\psdots[dotsize=0.4,linecolor=red,dotstyle=otimes](0.2,-0.0019192393)
\psdots[dotsize=0.4,linecolor=red,dotstyle=otimes](11.0,-0.0019192393)
\rput(5.793906,2.7080808){\LARGE $(\rho, \bar{\rho})$}
\rput{45.0}(1.0425268,-0.052283354){\rput(0.5515625,1.2080808){\LARGE $\bar{\rho}=-1$}}
\rput{-45.0}(2.343525,7.847369){\rput(10.611563,1.0680808){\LARGE $\rho=1$}}
\rput{-45.0}(0.9090884,3.2593324){\rput(4.356094,0.5080808){\LARGE $\rho=0$}}
\rput{45.0}(2.2228932,-4.66194){\rput(6.706094,0.32808077){\LARGE $\bar{\rho}=0$}}
\end{pspicture} 
}

\caption{The bulk channel $\rho$ coordinates defined in \cite{Lauria:2017}. The defect is a sphere and crosses the plane at the location of the two red dots. The operators are drawn in the time-like configuration that corresponds to $0<z<1/\bz<1$, that is the region of integration in eq.~\eqref{eq:inversionformula}.}
\label{fig:rhobulk}
\end{figure}

Following \cite{Lauria:2017}, let us define the following change of variables:
\beq
\rho=\frac{1-\sqrt{z}}{1+\sqrt{z}}~, \qquad \bar{\rho}=\frac{1-\sqrt{\bz}}{1+\sqrt{\bz}}~.
\eeq
The configuration of the operators in the $\rho$ coordinates is depicted in fig.~\ref{fig:rhobulk}. The defect is now spherical, and crosses the $(\rho,\bar{\rho})$ plane in two points. The configuration is of course very similar to the one customarily used to describe a four-point function of local operators \cite{Hogervorst:2013sma}. 
In the region $0<\rho,\bar{\rho}<1$ the bulk-channel OPE has the following structure, for a scaling operator of dimension $\Delta$ and $SL(2,\mathbb{R})$ spin $J$:
\beq
g(\rho,\rhob) \sim \rhob^{\frac{\Delta-J}{2}}\rho^{\frac{\Delta+J}{2}}~.
\label{sketchybulk}
\eeq
The integral in the Lorentzian inversion formula \eqref{eq:inversionformula} extends over the region $(0<\rho<1, -1<\rhob<0)$, corresponding to the configuration of the operators drawn in figure \ref{fig:rhobulk}. The bulk channel OPE in this region still has the form \eqref{sketchybulk}, and is only modified by the phases picked up going past the singularity at $\rhob=0$. Since the OPE is absolutely convergent in the region $\rho,\bar{\rho}<1$, it still converges after the continuation.

The $w\to 0$ limit at fixed $r$ is the limit in which the operators reach the upper and lower corners of the causal diamond of the defect.

\bibliography{./auxi/bibliography}
\bibliographystyle{./auxi/JHEP}

\end{document}